\documentclass[aps,prd,nofootinbib,superscriptaddress,twocolumn]{revtex4-2}
\usepackage{amsmath}
\usepackage{amsthm}
\usepackage{amsfonts}
\usepackage{natbib}
\usepackage{physics}
\usepackage{graphicx}
\usepackage[usenames,dvipsnames]{color}
\usepackage{orcidlink}

\allowdisplaybreaks

\def\lsim{~\rlap{$<$}{\lower 1.0ex\hbox{$\sim$}}}
\def\bsim{~\rlap{$>$}{\lower 1.0ex\hbox{$\sim$}}}

\def\hmmpc{\ {\rm {\it h}Mpc^{-1}}}

\def\hmsun{\ {\rm M_\odot/{\it h}}}

\def\hhhmpc{\ {\it h}^{3}{\rm Mpc}^{-3}}

\def\hmmpc{\ {\rm {\it h}Mpc^{-1}}}

\def\hhhgpc{\ {\rm {\it h}^{-3}Gpc^{3}}}

\def\apjl{Astrophys.\ J.\ Lett.}
\def\mnras{Mon.\ Not.\ R.\ Astron.\ Soc.}

\def\aap{Astron.\ Astrophys.}
\def\apj{Astrophys.\ J.}

\def\apjs{Astrophys.\ J. Supp.}
\def\prl{Phys.\ Rev.\ Lett.}
\def\prd{Phys.\ Rev.\ D}

\def\physrep{Phys. Rep.}

\def\jcap{{JCAP\ }}

\newcommand{\be}{\begin{equation}}
\newcommand{\ee}{\end{equation}}

\let\ln\relax
\DeclareMathOperator{\ln}{ln}

\def\mathbi#1{\textbf{\em #1}}

\def\dd{\mathrm{d}}

\def\nvh{\hat{\mathbi{n}}}

\def\vk{\mathbi{k}}

\def\vq{\mathbi{q}}
\def\vr{\mathbi{r}}

\def\vv{\mathbi{v}}

\def\vx{\mathbi{x}}
\def\vvx{\mathbi{X}}

\def\cH{\mathcal{H}}

\def\cH{\mathcal{H}}

\def\pde{p_\text{DE}}
\def\rde{\rho_\text{DE}}

\begin{document}
	
	\title{Sound Mode and Scale-Dependent Growth in Two-Fluid Dynamical Dark Energy}
	
	\author{Frans van Die\,\orcidlink{0009-0002-6134-5881}}
	\email{frans.van@campus.technion.ac.il}
	\affiliation{Physics department, Technion, Haifa 3200003, Israel}
	
	\author{Vincent Desjacques\,\orcidlink{0000-0003-2062-8172}}
	\email{dvince@physics.technion.ac.il}
	\affiliation{Physics department, Technion, Haifa 3200003, Israel}
	
	\begin{abstract}
		We investigate the effects of dynamical dark energy (DDE) on the growth of cosmic structure using a two-fluid model. 
		This framework allows the dark energy equation of state to smoothly cross the phantom divide, in agreement with recent DESI results. 
		In this effective description, DDE supports propagating perturbations that behave like sound waves. 
		These perturbations induce a scale dependence in the growth of matter fluctuations and in halo bias,
		which can be exploited to test the dynamical nature of dark energy at the level of its fluctuations.
		For cluster-sized halos, the amplitude of the scale-dependent halo bias is comparable to that produced by massless neutrinos in $\Lambda$CDM.
		Using a Fisher forecast for a multi-tracer analysis of the power spectrum (P) and bispectrum (B) of galaxy number counts, we find that bispectrum information is essential to detect the scale dependence induced by the DDE sound mode.
		For a survey of volume $V\sim 10\hhhgpc$ at redshift $z=0.5 - 1$, a two-tracer P+B analysis could detect this scale dependence 
		if the sound speeds of the dark energy fluids are in the range $c_s^2\sim 10^{-2} - 10^{-4}$. 
		Lower sound speeds cause halos to experience a gravitational drag force through the excitations of sound waves. 
		This effect impacts measurements of the growth rate inferred from cluster-sized halos at the 10\% level if one of the fluids has a very low sound speed $c_s^2\sim 10^{-5}$. 
		Larger sound speeds $c_s^2 > 10^{-2}$ could be probed with optimal weighting schemes that reduce shot noise and increase the effective bias.
	\end{abstract}
	
	\maketitle
	
	\section{Introduction}
	\label{sec:introduction}
	
	Dynamical dark energy (DDE) models generalize the standard cosmological constant scenario by introducing a time-dependent equation of state (EOS) parameter $w=\pde/\rde$, where $\pde$ and $\rde$ are the pressure and the energy density of the dark energy (DE) component, respectively. While the cosmological constant corresponds to $w=-1$, DDE models allow $w$ to evolve with cosmic time and thus capture a broader class of late-time cosmic acceleration scenarios. 
	Among these models, quintessence \cite{caldwell/etal:1998, Zlatev:1998tr} refers to scenarios where DE arises from a scalar field with a canonical kinetic term, characterized by a sound speed $c_s=1$. In contrast, $k$-essence models exhibit non-canonical kinetic terms with $c_s<1$ \cite{peebles/ratra:1988,frieman/etal:1995,caldwell/etal:1998,erickson/etal:2002}.
	Specific examples of $k$-essence include an evolving scalar field \cite{Armendariz-Picon:1999hyi, Armendariz-Picon:2000nqq}, the ghost condensate model \cite{arkani-hamed/etal:2004}, and dusty dark energy \cite{lim/etal:2010,Sebastiani:2016ras} (see \cite{kunz:2012} for a concise review of the fluid description of these models).
	DDE models may also be reformulated as interacting dark energy (IDE) models, where energy exchange occurs between DE and dark matter (DM) components (DDE and IDE can be mathematically equivalent under certain conditions, see \cite{Tamayo:2025wiy}).
	These models can have significant consequences for cosmic structure formation and dynamics \cite[e.g.][]{Putter, Linton:2017ged,hassani/etal:2019,hassani/etal:2022, Ishiyama:2025bbd, Neomenko:2024aix}, with some variants predicting clustering or interactions in the dark sector \cite{Yang:2025gaz,DAmico:2020tty, Wu:2025vrl}. Most observational hints of DDE (or any departure from a cosmological constant) come from the late-time (low-redshift) Universe, including those interpreted as evidence for IDE \cite{Zhai:2025hfi}. 
	
	Recent results from the DESI survey have sparked renewed interest in DDE by suggesting a time-dependent EOS with a maximum energy density at a redshift $z\sim 0.5$ \cite{DESI:2024hhd,DESI:2024mwx,DESI:2025zgx,DESI:2025gwf}. 
	Follow-up studies have pointed out the role of tensions among different datasets \cite{Wang:2025bkk,Teixeira:2025czm,Capozziello:2025qmh} and the influence of priors \cite{Toomey:2025xyo} on the DESI results. Although the measurement largely relies on the standard Chevallier–Polarski–Linder (CPL) parameterization \cite{Chevallier:2000qy,Linder:2002et}, which may be insufficient to capture the full complexity of evolving DE models \cite{Nesseris:2025lke,Ormondroyd:2025exu,Giare:2025pzu,Shlivko:2025fgv,Artola:2025zzb,Toomey:2025xyo,DESI:2025fii}, several independent analyses based on alternative parameterizations or non-parametric methods have reached similar conclusions \cite{Gialamas:2024lyw,Keeley:2025stf,Roy:2024kni,Berti:2025phi,Bansal:2025ipo,Shajib:2025tpd,gu/etal:2025, Kessler:2026dbi}.
	Analyses of other cosmological datasets, such as KiDS \cite{Reischke:2025hrt}, BOSS\cite{chen/ivanov/etal:2024,chen/ivano:2025,silva/nunes:2025} and DES \cite{Lopez-Hernandez:2025lbj,DES:2025bxy,DES:2026aht,DES:2026jmi}, reach conflicting conclusions with regard to the evidence for DDE. 
	Alternative explanations, such as unaccounted local systematics, calibration issues, or the presence of large cosmic voids, inhomogeneities in general, or modified recombination have been proposed to explain the data \cite{Colgain:2024mtg,Dinda:2025hiu,Gialamas:2024lyw,Moffat:2025sik,Mirpoorian:2025rfp,Camarena:2025upt,Ginat:2026fpo,Banik:2026imu,Battye:2026tsu, Macpherson:2026bzj}.
	Future measurements focusing on gravitational wave standard sirens \cite{Wang:2020xwn,Santos:2025wiv} or Lyman-$\alpha$ forest data \cite{Garza:2026kys} offer potential avenues for an independent confirmation of the DESI results.
	
	Assuming an evolving dark energy, theoretical challenges remain to explain the DESI measurements, which indicate that a thawing DE model up-crossing the so-called phantom divide ($w=-1$) fits the data best \cite{Gu:2025xie}. In particular, single-field models cannot cross $w=-1$ without instabilities. Crossing requires either non-minimal couplings or multiple degrees of freedom, as in quintom models, which combine quintessence and phantom components to yield an effective phantom crossing \cite{Linder:2002et,Fang:2008sn,Amendola:2016saw,Kunz:2006wc}. However, such models often challenge energy conditions and quantum stability \cite{Amendola:2016saw}. 
	Putting these model-dependent considerations aside, DDE has internal degrees of freedom, unlike the cosmological constant, which can be excited by fluctuations in the energy density. New propagating DE degrees of freedom imply the existence of a propagating scalar mode with an effective sound speed, i.e. a sound mode. 
	The existence of this sound mode is generic in standard realizations of DDE (e.g. quintessence, $k$-essence) \cite{Perrotta:2002sw,dePutter:2007ny,dePutter:2010vy,Calabrese:2010uf} or in the effective field theory (EFT) of dark energy \cite{Gubitosi:2012hu,Gleyzes:2013ooa}.
	The response of DDE to gravitational disturbances induces a scale-dependent modification of the growth of matter fluctuations, which in turn imprints a characteristic scale dependence in the bias of large-scale structure tracers. 
	
	In this paper, we explore these effects and assess the feasibility of their detection within the framework of an effective, two-fluid DDE model \cite{hu:2005}, which allows a smooth transition at the phantom crossing and can be made consistent with the DESI measurements. This two-fluid DDE model is summarized in Section \S\ref{sec:model}. 
	The characteristics of the scale-dependent growth and bias and their observational detection are explored in Section \S\ref{sec:scalekbias}. Furthermore, if DDE supports a sound mode, it acts as a dissipative medium, through which DM halos experience dynamical friction (DF). This effect is discussed in Section \S\ref{sec:df}. Our findings are summarized in Section \S\ref{sec:conclusions}.
	The two-fluid approach adopted here should be regarded as an effective description of generic DDE properties rather than as a fundamental model. Our analysis therefore serves as a proof of concept, illustrating how scale dependence and velocity statistics could probe the sound mode of a DDE consistent with DESI data in a largely model-independent manner.

	\section{Dynamical Dark energy model}
	\label{sec:model}
	
	\subsection{Equation of state}
	
	DDE can be parameterized phenomenologically using, for example, a time-varying EOS
	\begin{align}
		w(a) = w_0 + w_a(1-a)
	\end{align}
	as in the CPL model \cite{Chevallier:2000qy,Linder:2002et}. 
	Here, $w_0$ and $w_a$ are constants and $a$ is the scale factor. 
	The properties of DE enter the cosmological measurements through their impact on cosmic distances. 
	Assuming a spatially flat background, the first Friedmann equation gives \cite{Lodha:2024kob}
	\begin{align}
		\frac{H^2}{H_0^2} &= \Omega_{m,0}a^{-3} + \Omega_{r,0}a^{-4} 
		+ \Omega_{\rm DE}(a), \label{eq:Hubble}
	\end{align}
	where $\Omega_{\rm DE} = (1-\Omega_{m,0} - \Omega_{r,0})f_{\rm DE}(a)$ describes the evolution of the DDE energy density
	while $f_{\rm DE}(z)$ is related to the equation-of-state parameter $w(a)$ through 
	\begin{align}
		w(a) = -1 - \frac{1}{3} \dv{\ln f_{\rm DE}}{\ln a} \;.
	\end{align}
	For the CPL parameterization, we have
	\begin{align}
		f_{\rm DE}(a) = a^{-3(1+w_0+w_a)}\, e^{-3w_a(1-a)} \;.
	\end{align}
	There are two distinct regions in the DDE phase space: the thawing ($w_a<0$) and freezing ($w_a>0$) regions \cite{Caldwell:2005tm}. 
	The freezing region, with $w_a > 0$, is disfavored by the latest DESI measurements.
	
	\subsection{A two-fluid approach}
	
	We follow \cite{hu:2005,Fang:2008sn} and model DDE with two effective fluids parameterized  by their EOS parameter $w_\pm$ and the propagation speed 
	$\hat c_\pm$ of sound waves (see \cite{kunz:2012, Mehrabi:2015hva} for concise discussions on the fluid description of modified DDE models and the role of the effective sound speed in matter clustering). 
	This two-fluid toy model resolves the energy divergence at phantom crossing, which arises in single-component models when the EOS parameter reaches $w=-1$ at which point the pressure perturbation becomes singular. By splitting DDE into two fluids with constant EOS, for instance, one can ensure that the total fluid crosses the phantom divide without encountering divergences.
	In what follows, we shall take $w_\pm = \bar P_\pm/\bar\rho_\pm$ to be time-independent, although this may be an unrealistic assumption for rolling scalar fields. 
	Concretely, we introduce
	\begin{equation}
		\label{eq:wpm}
		w_\pm = \bar w \pm \epsilon_n,
	\end{equation} which gives the relative energy density contributions
	\begin{equation}
		\rho_\pm(a) = \frac{1}{2}\rho_t(a) \big(1\pm \delta(a)\big) \;.
	\end{equation}
	Here, $\rho_t(a)$ is the total DDE energy density while
	\begin{equation}
		\delta(a) = \delta_n + \frac{(1-\delta_n^2)\big(1-(a/a_n)^{6\epsilon_n}\big)}{1+\delta_n+(1-\delta_n)(a/a_n)^{6\epsilon_n}}\;.
	\end{equation}
	The DE energy ratio $\delta_n\equiv\delta(a_n)$ is defined at the reference epoch $a_n$, which serves as the calibration point of the two-fluid parametrization. The choice of $a_n$ fixes where the CPL EOS matches the effective EOS of the two-fluid dark energy model. Changing $a_n$ requires a consistent refit of the remaining model parameters. Once $a_n$ is set, the target DDE EOS $w_0+(1-a) w_a$ together with a choice of $\epsilon_n$ determines the remaining model parameters $\bar\omega$ 
	and $\delta_n$ through the relations
	\begin{align}
		\label{eq:wb}
		\bar w &= w_0 + \big(1-a_n\big) w_a - \delta_n\,\epsilon_n \\
		\label{eq:dn}
		\delta_n &= \sqrt{1 - \frac{w_a a_n}{3\epsilon_n^2}} \;.
	\end{align}
	The total EOS parameter is now given by
	\begin{equation}\label{eq:wt}
		w(a) = \bar w + \delta(a) \epsilon_n \;.
	\end{equation}
	Since $w_\pm$ are constant, the adiabatic sound speeds are given by $c_{a,\pm}^2 = \dot{\bar P}_\pm/\dot{\bar\rho}_\pm = w_\pm <0$, but these are not the propagation 
	speeds $\hat c_\pm$ of the sound waves. 
	
	To ensure $\hat c_\pm^2>0$, we must allow for non-adiabatic degrees of freedom ($\Gamma_\pm\ne 0$) and avoid the relation $\hat c_\pm^2 = c_{a,\pm}^2 < 0$, 
	which would lead to gradient instabilities. 
	For this purpose, we move to Fourier space and define the propagation speed squared $\hat c_\pm^2$ of sound waves in each of the DE fluid rest frame through \cite{Amendola:2016saw}
	\begin{equation}
		\label{eq:delhatP}
		\frac{\delta\hat P_\pm}{\bar \rho_\pm} = c_{a,\pm}^2 \hat\delta_\pm + w_\pm \Gamma_\pm \equiv \hat c_\pm^2(k) \hat\delta_\pm\;,
	\end{equation}
	where $k$ is the comoving wavenumber and fluid rest frame quantities are denoted with a circumflex. 
	Here, $\delta\hat P_\pm$ and $\hat\delta_\pm$ are the Fourier modes of the pressure (isotropic stress) 
	and fractional energy density perturbations, while $c_{a,\pm}^2 \hat\delta$ and $w_\pm\Gamma_\pm$ are the adiabatic and non-adiabatic contributions to $\delta\hat P_\pm$.
	The quantity $\Gamma_\pm$ is a gauge-independent perturbation related to the entropy flux \cite{kodama/sasaki:1984}.
	In a general gauge, we thus have \cite{Amendola:2016saw}
	\begin{equation}
		\label{eq:delP}
		\frac{\delta P_\pm}{\bar \rho_\pm} = c_{a,\pm}^2\, \delta_\pm + w_\pm \Gamma_\pm
	\end{equation}
	where $\delta P_\pm$ and $\delta_\pm$ are gauge-dependent pressure and fractional density perturbations. 
	
	The fluid rest-frame fractional energy densities $\hat\delta_\pm$ can be expressed in terms of the fractional energy density perturbation $\delta_\pm$ and the velocity perturbation 
	$\theta_\pm$ (written here as $\theta_\pm = k V_\pm$) in any gauge through
	\begin{equation}
		\label{eq:hatdelta}
		\hat\delta_\pm = \delta_\pm + 3\cH\big(1+w_\pm\big) \frac{V_\pm}{k}
	\end{equation}
	where $\cH=aH$ is the conformal Hubble rate.
	Therefore, on combining Eqs.~\eqref{eq:delhatP}--\eqref{eq:hatdelta} and using $c_{a,\pm}^2 = w_\pm$, the pressure perturbation reads
	\begin{equation}
		\frac{\delta P_\pm}{\bar\rho_\pm} = \hat c_\pm^2 \,\delta_\pm + 3\cH\big(1+w_\pm\big) \big(\hat c_\pm^2 - w_\pm\big) \frac{V_\pm}{k} \;. 
	\end{equation}
	In the absence of anisotropic stress, this prescription closes the system of equations (conservations laws and Einstein equations). 
	
	\subsection{Parameter choice}
	
	In this work, we set $a_n=0.75$, because it places the matching point around the phantom crossing where the effects on observables are quite pronounced, and we consider two different (two-fluid) DDE models: 
	(i) a "Phantom" (thawing) model with $w_0 = -0.752$ and $w_a = -0.86$ (consistent with the DESI+CMB+DESY5 measurements reported in \cite{DESI:2025zgx}) and  
	(ii) a "Quintom" (freezing) model with $w_0 = -1.15$ and $w_a = 0.5$.  
	Although the Quintom model does not fit the DESI data, as it crosses the phantom divide in the opposite direction, it provides a useful benchmark for comparison 
	with the Phantom model.
	
	As discussed above, these models have three free parameters ($\bar{w}$, $\epsilon_n$, and $\delta_n$) in addition to the sound speeds $\hat c_\pm$,
	whereas the CPL parameterization has only two ($w_0$ and $w_a$). 
	Therefore, one of the model parameters can be used to close the parameterization and fix the remaining freedom in the two-fluid decomposition.
	We set $\epsilon_n$ such that the two-fluid models satisfy the conditions spelled out in \cite{hu:2005}.
	First of all, they should match a reasonably large range of EOS parameters, i.e. $-1<w_a<1$, 
	and the limit $w_a \rightarrow 0$ should correspond to a single fluid DE model. 
	These two conditions can be realized if $\epsilon_n$ is a power-law function of $\abs{w_a}$ of the form
	\begin{equation}
		\epsilon_n = \left(\frac{a_n}{3}\right)^{1/2}\, |w_a|^{(1-p)/2}
		\label{eq:en}
	\end{equation}
	for $\abs{w_a}<1$.
	The mitigation of higher-order derivatives of $w(a)$ such as $w_{aa} = {\rm d}^2 w/{\rm d}a^2$ etc. -- which do not appear in the CPL parameterization -- 
	provides a third condition, which can be enforced with $p = 0.1$. 
	The various model parameters are summarized in Table \ref{tab:hu_params}. 
	Once $(a_n,p,w_0,w_a)$ are specified, all the other quantities follow. 
	The table illustrates that, while the Phantom model is a combination of two phantom fluids ($w_\pm < -1$), a "microscopic" description of the Quintom model 
	would involve a $k$-essence field with $w_+>-1$. 
	
	\begin{figure}[h!]
		\centering
		\includegraphics[width=0.48\textwidth]{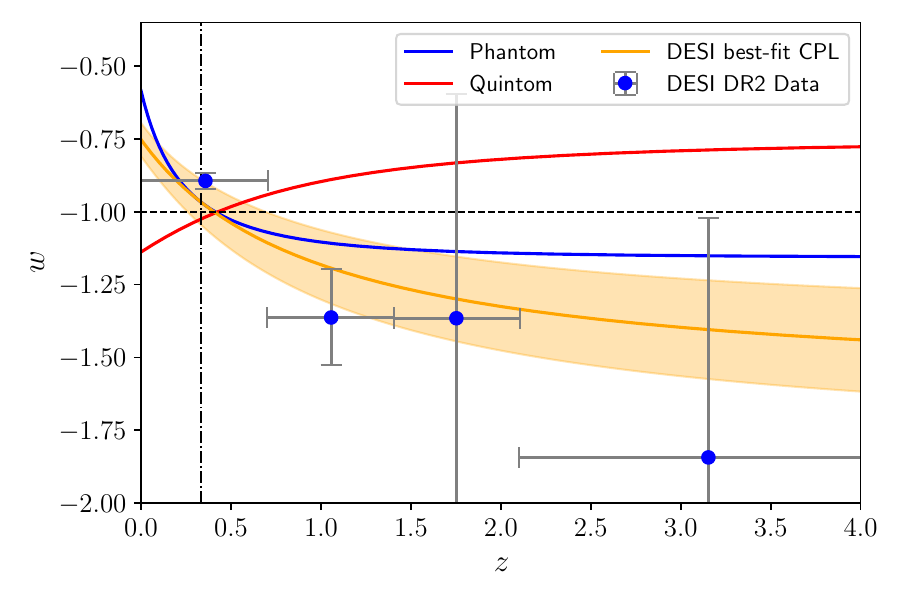}
		\caption{Redshift evolution of the total EOS parameter $w_t(z)$, following Eq. \eqref{eq:wt}, with the parameters of Table \ref{tab:hu_params}. 
			The dashed line represents the phantom divide at $w=-1$, while the dashed-dotted line represents the reference epoch $a=a_n$.}
		\label{fig:totw}
	\end{figure}
	
\begin{figure}[h!]
		\centering
		\includegraphics[width=0.48\textwidth]{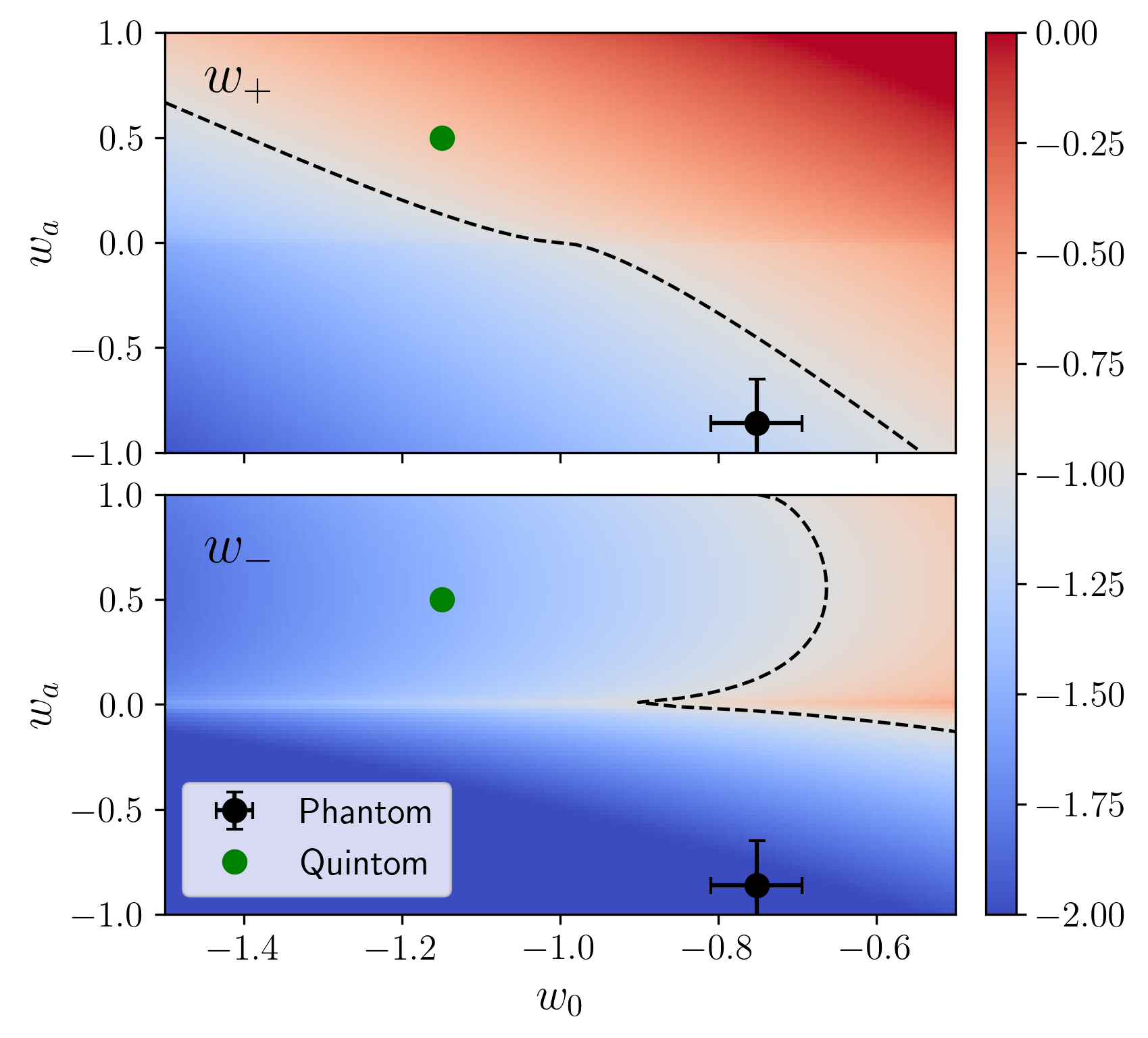}
		\caption{The dependence of the EOS parameters $(w_+,w_-)$ on the CPL parameters $(w_0,w_a)$ using the two-fluid approach discussed in the text.
			The color scale represents either $w_+$ (top panel) or $w_-$ (bottom panel).
			In each panel, the two circles indicate the CPL parameter values of the Phantom and Quintom model considered here.
			For the Phantom model, we have added the uncertainties on $(w_0,w_a)$ reported in \cite{DESI:2025zgx} since, in this case, our choice of $(w_0,w_a)$ 
			is motivated by the DESI DR2 results.}
		\label{fig:wowawpwm}
	\end{figure}

	Fig.~\ref{fig:totw} shows the redshift evolution of the (total) EOS parameter $w(z)$ in the Phantom and Quintom models calibrated to cross the 
	phantom divide ($w=-1$) around $a=a_n$. For comparison, the data points show the DESI DR2 measurements while the shaded area indicates the DESI best-fit CPL model. 
	The DESI redshift binned results suggest $w_a<0$, which cannot be achieved unless the two DE fluids are phantom. 
	However, even in the case of two phantom fluids, the total EOS parameter can shift to the quintessence regime. 
	To visualize this better, Fig. \ref{fig:wowawpwm} shows how $w_\pm$ vary as a function of $(w_0,w_a)$, and which part of the parameter space correspond to phantom 
	or quintessence fluids.
	
	\begin{table}[h!]
		\centering
		\caption{Summary of parameter choices in the two-fluid, Phantom and Quintom DE models.
			The basic parameters $(a_n,p,w_0,w_a,\hat c_\pm,\Omega_{m0},h)$, with $H_0 = 100\,h\,{\rm km\,s^{-1}\,Mpc^{-1}}$, are listed in the upper half of the table.
			They are chosen based on physical arguments and observational constraints, as explained in the text.
			The lower half includes the parameters inferred from Eqs.~(\ref{eq:wpm}, \ref{eq:wb}, \ref{eq:dn}, \ref{eq:en}).}
		\label{tab:hu_params}
		\begin{tabular}{|c|c|c|c|}
			\hline
			Parameters & Meaning & Phantom & Quintom \\
			\hline\hline
			$a_n$ &  Reference epoch & 0.75 & 0.75 \\
			\hline
			$p$ &  Power-law index & 0.1 & 0.1 \\
			\hline
			$w_0$ &  CPL present EOS & -0.752 & -1.15 \\
			\hline
			$w_a$ &  CPL EOS evolution & -0.86 & 0.5 \\
			\hline
			$\Omega_{m0}$ &  Present matter density & 0.319 & 0.315 \\
			\hline
			$h$ &  Reduced Hubble rate & 0.6674 & 0.674 \\
			\hline\hline
			$\delta_n$ &  DE energy ratio at $a_n$ & 1.40 & 0.26 \\
			\hline
			$\epsilon_n$ &  EOS splitting & 0.47 & 0.37 \\
			\hline
			$\bar{w}$ &  Mean EOS & -1.63 & -1.12 \\
			\hline
			$w_-$ &  Fluid 1 EOS & -2.10 & -1.49 \\
			\hline
			$w_+$ &  Fluid 2 EOS & -1.16 & -0.75 \\
			\hline
			$\Omega_{-0}$ &  Present density fluid 1 & -0.42 & 0.36 \\
			\hline
			$\Omega_{+0}$ &  Present density fluid 2 & 1.10 & 0.33 \\
			\hline
		\end{tabular}
	\end{table}
	
\begin{figure}[h!]
		\centering
		\includegraphics[width=0.48\textwidth]{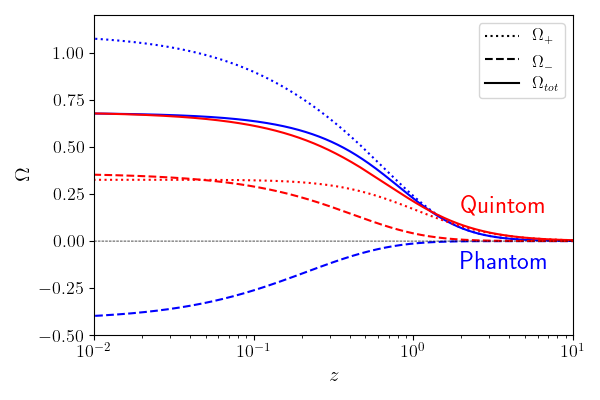}
		\caption{Redshift evolution of the dimensionless energy densities of both DE fluids, as well as the sum in both the Phantom and the Quintom model. 
			The black dashed line indicates $\Omega = 0$. 
			Although the total energy density remains positive, the $w_-$ fluid in the Phantom model develops a negative energy density at late times 
			(which is not directly measurable in the effective framework adopted here).}
		\label{fig:omegaevol}
	\end{figure}

	As for the cosmologies used in each of the different models, the Phantom model relies on the best-fit DESI+CMB+DESY5 values for $\Omega_{m0}$ and $H_0$ \cite{DESI:2025zgx}, 
	while for the Quintom model, we use the cosmological parameter results from Planck \cite{Planck:2018vyg}. 
	A summary of this is added to Table \ref{tab:hu_params}. 
	In Fig.~\ref{fig:omegaevol}, we show the evolution of the dimensionless energy parameters $\Omega_-$, $\Omega_+$ of the DE fluids and their sum, for both models. 
	Fig.~\ref{fig:omegaevol} and Table \ref{tab:hu_params} show that, in the Phantom model, $\rho_-<0$ once DE becomes dynamically relevant. 
	However, this does not imply a violation of the weak energy condition because, in this effective two-fluid description, only the behavior of the total energy density matters.
	The latter is well-behaved since $\rho_++\rho_- > 0$ at all times (see also \cite{Caldwell:2025inn} for a related discussion).
	
	Finally, CMB measurements set limits on the value of the speed of sound of DDE (treated as a single effective fluid) through the ISW effect \cite{hu/scranton:2004,bean/dore:2004}.
	CMB alone tentatively suggests $\hat c_s^2 < 0.04$ (at the 1-$\sigma$ level) on these large scales \cite{bean/dore:2004}, although this claim is debated \cite{Corasaniti:2005pq}. 
	In what follows, we will assume that $\hat c_\pm$ are dependent and loosely bounded by $\hat c_\pm^2 \gtrsim 10^{-5}$ to ensure that DE does not cluster significantly at 
	small scales (see also Section~\S\ref{sec:df}).
	
	\section{Scale-dependence}
	\label{sec:scalekbias}
	
	The scale-independent growth of matter fluctuations predicted in the $\Lambda$CDM model is violated in DDE models. 
	A fluid with sound speed $c_s<1$ introduces a scale-dependence in the growth rate of matter density perturbations relative to a quintessence field with $c_s=1$ 
	(among other effects, e.g. \cite{Batista_2017,Basse_2011,Dent:2009wi}).
	This scale dependence manifests itself on scales of order the DE sound horizon, where pressure gradients are significant, that is, for $k\gtrsim k_{\pm}$ 
	where $k_{\pm} = \cH/\hat c_\pm$ is the Jeans wavenumber.
	The resulting scale-dependent growth leads to a scale-dependent bias \cite{parfrey/hui:2008,parfrey/etal:2011,chiang/etal:2016} at redshift $z\lesssim z_\text{DE}$, after the 
	epoch $z_\text{DE}$ at which DDE becomes dynamically relevant in the background evolution.
	
	\subsection{Scale-dependent growth}
	
	To calculate the magnitude of this effect in the two-fluid DDE model considered here, we will assume that both fluids have negligible anisotropic stress 
	(i.e. $\Pi_\pm\approx 0$) so that the gravitational (Bardeen) potentials satisfy $\Phi=\Psi$ (e.g. \cite{Anselmi:2011ef}).
	Furthermore, we shall ignore the metric perturbations induced by the components of the stress-energy tensor other than matter and DE.
	On the sub-horizon scales $k/\cH\gg 1$ of interest here, the equation governing the growth of linear matter perturbation reduces to 
	\begin{equation}
		\label{eq:Dmcons}
		\frac{{\rm d}^2\delta_m}{{\rm d}\ln a^2}+\bigg(1+ \dv{\ln\cH}{\ln a}\bigg)\dv{\delta_m}{\ln a}+ \frac{k^2}{\cH^2}\Psi = 0 
	\end{equation}
	in the conformal Newtonian gauge.
	Here $\delta_m=\delta_m(\eta,\vk)$ are the (Fourier mode of the) matter density fluctuations, $\eta$ is the conformal time and $k$ is the comoving wavenumber.
	The gravitational potential $\Psi(\eta,\vk)$ is determined by Poisson's equation
	\begin{equation}
		k^2 \Psi = -\frac{3}{2}\cH^2 \Big(\Omega_m\delta_m + \Omega_+\delta_+ + \Omega_-\delta_-\Big) \;.
	\end{equation}
	Neglecting again time derivatives of $\Psi$, the linearized energy and momentum conservation equations of the individual DE fluids read
	\begin{align}
		\delta_\pm' + 3 \cH\big(\hat c_\pm^2 - w_\pm\big) \delta_{\pm}+\big(1+w_\pm\big) kV_\pm &=0 \label{eq:Econs0} \\
		V_\pm' + \cH\big(1-3\hat c_\pm^2\big)V_\pm- k\frac{\hat c_\pm^2 \delta_\pm}{1+w_\pm} - k \Psi &= 0 \label{eq:Pcons0} \;,
	\end{align}
	where $\delta_\pm=\delta_\pm(\eta,\vk)$ and $V_\pm=V_\pm(\eta,\vk)$ are the energy density and velocity perturbation of the DE fluids and a prime denotes a derivative w.r.t. $\eta$.
	For constant DE equations of state and sound speeds, the second-order ODEs for the energy density perturbations $\delta_\pm$ take the form
	\begin{align}
		\label{eq:growthd}
		\frac{{\rm d}^2\delta_\pm}{{\rm d}\ln a^2}&+\bigg(1-3 w_\pm + \dv{\ln\cH}{\ln a}\bigg)\dv{\delta_\pm}{\ln a}+\bigg(1-3\hat c_\pm^2 \\
		& + \dv{\ln\cH}{\ln a}+\frac{k^2}{\cH^2}\hat c_\pm^2\bigg)\delta_\pm 
		+\frac{k^2}{\cH^2}\big(1+w_\pm\big) \Psi = 0 \nonumber
	\end{align}
	on sub-horizon scales $k\gg \cH$ (where $\cH\sim 3\times 10^{-4}\hmmpc$ for $0<z<2$).
	This makes clear that the $k$-dependence of $\delta_\pm$ is significant only on scales $k\sim k_\pm$. On sub-Jeans scales $k\gg k_\pm$, the overdensity $\delta_\pm$ drops to zero 
	as the perturbations become pressure supported. 
	In this limit, the growth rate of matter fluctuations is determined purely by the background cosmology. 
	The only deviation from a $\Lambda$CDM cosmology arises from the presence of two fluids rather than a cosmological constant.
	This is the regime considered in \cite{hu:2005}.  
	For $k\ll k_\pm$, the DE fluids cluster roughly scale independently, like non-relativistic matter, at least as long as $(1+w_\pm)$ is not too small and does not make the gravitational 
	source term negligible (see e.g.~\cite{Batista_2017,Basse_2011,Dent:2009wi}).
	
	\begin{figure}
		\centering
		\includegraphics[width=0.48\textwidth]{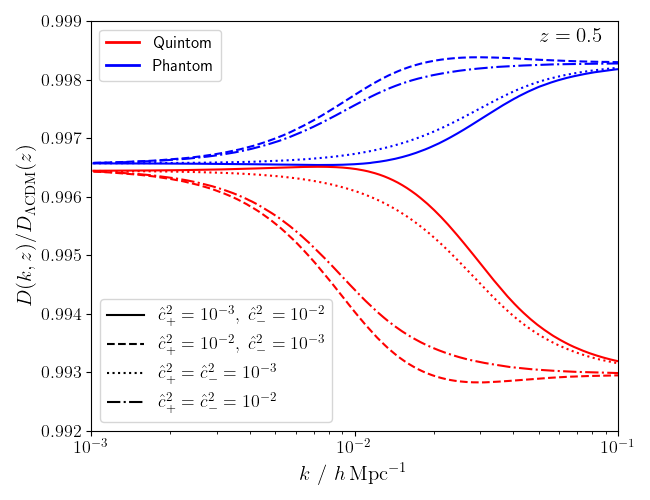}
		\caption{The scale-dependent, linear growth rate $D(k,z)$ of matter fluctuations, normalized to the $\Lambda$CDM solution, is shown at redshift $z=0.5$ 
			for several choices of $\hat c_\pm^2$.}
		\label{fig:growth}
	\end{figure}
	
	At fixed comoving wavenumber $k_\ast$, the relative importance of the two dark-energy fluids in sourcing the gravitational potential is determined by the combination $\Omega_\pm(z)\,\delta_\pm(k_\ast,z)$ entering the Poisson equation. 
	To determine whether we get a growth suppression or a growth enhancement on scales $k_\ast \ll k_\pm$ where pressure support is negligible, the quantity of interest is $\Omega_+\delta_+$
	since it is the $w_+$ fluid that constitutes the dominant contribution to the DE energy density in both models (down to low redshift, as is evident from Fig. \ref{fig:omegaevol}). 
	Noting furthermore that $\delta_\pm \propto (1+w_\pm)$, and $\Omega_+>0$ for both the Phantom and Quintom models, we understand why at large scales growth is suppressed in the Phantom model 
	($w_+<-1$), while it is enhanced in the Quintom model $(w_+>-1$).
	
	Fig.~\ref{fig:growth} shows the linear growth factor $D(k,z)$ at redshift $z=0.5$ normalized to the $\Lambda$CDM prediction, assuming the matter dominated growing mode solution and 
	the adiabatic relation $\delta_{\pm}(z_i)=(1+w_\pm)\delta_{m}(z_i)$ at the initial redshift $z_i=199$ (where DE is dynamically irrelevant). Even in the case where one chooses different initial conditions, e.g. $\delta_\pm'=0$, the results are not altered in a significant way. 
	At low wavenumber, the similarity of the linear growth rates in the Quintom and Phantom models is fortuitous.
	
	The values of $\hat c_\pm^2 = 10^{-3} - 10^{-2}$ are chosen such that $\cH/\hat c_\pm$ brackets the range of wavenumbers below which horizon-scale effects become important, and above 
	which the sound horizon falls in the mildly nonlinear regime. The scale-dependence increases with decreasing redshift. 
	For $z=0.5$ shown here, it builds up mainly around $k\sim k_+$ because $\Omega_+$ is a few times larger than $\Omega_-$ in both models (again, see Fig.~\ref{fig:omegaevol}). The amplitude
	of the scale-dependence depends mainly on the gravitational potential and, therefore, only weakly on the value of the sound speeds. In Section \ref{sec:scaledep}, we expand further on the redshift dependence of these processes.

	\subsection{Scale-dependent bias}
	\label{sec:scaledep}
	
	To begin, we focus on the scale-dependence in the linear bias $b_1$ before discussing higher-order bias.
	
	\subsubsection{Separate universe approach}
	
	The separate universe (SU) technique \cite{sirko:2005,gnedin/etal:2011,li/etal:2014,wagner/etal:2015,lazeyras/etal:2016} can provide accurate predictions for the scale-dependent bias
	induced by the scale-dependent growth of matter fluctuations. 
	Following \cite{Loverde:2014,chiang/etal:2016}, the scale-dependent linear bias of DM halos at the collapse redshift $z$ is given by
	\begin{equation}
		\label{eq:biaskSU}
		b_1(z,M,k) = 1 - b_1^L(z,M)\,\dv{\delta_c}{\delta_m}\
	\end{equation}
	where the critical overdensity for collapse $\delta_c=\delta_c(z,k)$ depends on the wavenumber $k$ of the long mode $\delta_m=\delta_m(z,k)$.
	Furthermore, the (scale-independent piece of the) linear Lagrangian bias is
	\begin{equation}
		b_1^L(z,M) = -\frac{1}{\sigma}\dv{\ln \bar n}{\nu}
	\end{equation}
	for a universal mass function that depends only on the peak height $\nu=\delta_c/\sigma$, where $\sigma$ is the variance of matter fluctuations on the small (halo) scale. 
	To calculate $({\rm d}\delta_c/{\rm d}\delta_m)(z,k) \equiv {\rm d}\delta_c(z,k)/{\rm d}\delta_m(z,k)$, 
	we use the spherical collapse to model the evolution of the (physical) radius $R$ of the collapsing DM halo, i.e.
	\begin{align}
		\frac{{\rm d}^2 R}{{\rm d}\ln a^2}&+\dv{\ln H}{\ln a}\dv{R}{\ln a}+\frac{GM}{H^2 R^2} \\
		&\quad =-\frac{1}{2}\sum_{\alpha=\pm}\Omega_\alpha\Big[(1+3w_\alpha)+(1+3 \hat c_\alpha^2)\delta_\alpha\Big] R \nonumber
	\end{align}
	upon substituting $\delta P_\pm = \hat c_\pm^2 \bar\rho_\pm \delta_\pm$ for the sub-horizon scales of interest. 
	We assume adiabatic growing-mode initial conditions at $a_i=10^{-3}$ (any isocurvature component will become sub-dominant at late time and is thus neglected).
	This choice fixes the initial value $(\delta_{mi},\delta_{+i},\delta_{-i})$ of the long mode in the matter dominated era, 
	and determines the initial values $R_i$ and $({\rm d}R/{\rm d}\ln a)_i$ through the relation
	\begin{equation}
		\left(\dv{R}{\ln a}\right)_i = \bar R_i \left(1-\frac{\delta_i+\delta_{mi}}{3}\right)^2 = \frac{R_i^2}{\bar R_i}
	\end{equation}
	where $\bar R_i=(3M/4\pi\bar\rho_m(a_i))^{1/3}$ and $\delta_i$ is the initial density contrast of the spherical perturbation. 
	The response ${\rm d}\delta_c/{\rm d}\delta_m$ to the long mode is computed by tuning $\delta_i$ such that the spherical perturbation collapses (i.e. $R=0$) at the same collapse redshift $z$.
	In practice, we consider two different long modes with $\delta_{mi}=\pm 0.1\delta_i$ to calculate ${\rm d}\delta_c/{\rm d}\delta_m$.
	Eq.~(\ref{eq:biaskSU}) is our first estimate of the scale-dependent bias induced by DDE.
	
	\subsubsection{Local Lagrangian bias approximation}
	
	Alternatively, we follow \cite{parfrey/hui:2008,parfrey/etal:2011} and assume that the linear order Lagrangian and
	matter density fields satisfy a scale-independent relation
	\begin{equation}
		\delta_h^L(z_i,\vk) = b_1^L(z_i,M)\, \delta_m(z_i,\vk) + \dots 
	\end{equation}
	to calculate the scale-dependent bias. This approximation provides a fast estimate of the late-time scale-dependent bias because the time evolution of the spherical overdensity 
	up to collapse is not needed.
	
	\begin{figure}
		\centering
		\includegraphics[width=0.48\textwidth]{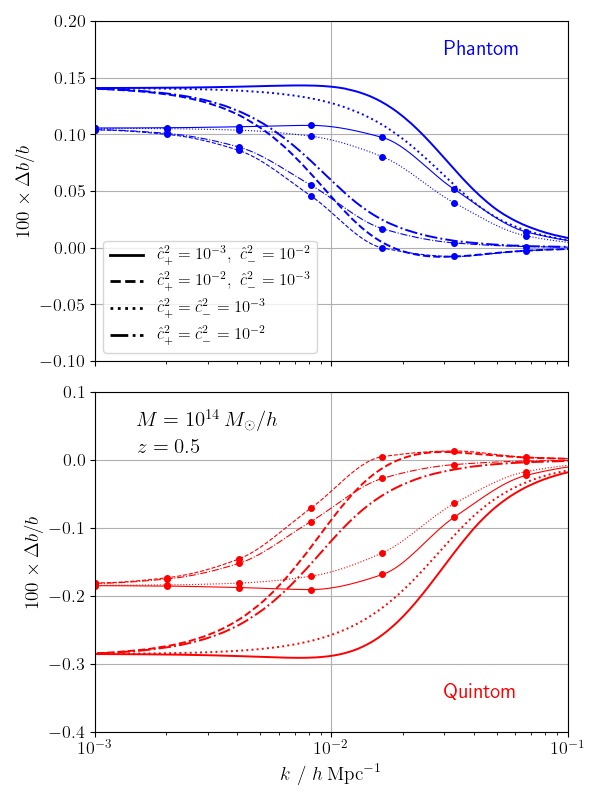}
		\caption{Scale-dependent bias computed in the separate universe approach (filled circles with a thin interpolating line) and in the local Lagrangian approximation (thick curves). 
			Results are shown at $z=0.5$ for four different choices of sound speeds assuming a DM halo mass $M=10^{14}\hmsun$.}
		\label{fig:biask_SUvsLL}
	\end{figure}
	
	Under this assumption, we relate the initial DM halo overdensity field $\delta_h^L(z_i,\vk)$ at redshift $z_i\gg z_\text{DE}$ to the evolved one at $z\lesssim z_\text{DE}$, $\delta_h(z,\vk)$, 
	through the time- and scale-dependence of the linear theory fluctuation:
	\begin{align}
		\delta_h^L(z_i,\vk) &\approx b_1^L(z_i,M)\left(\frac{D(z_i,k_\downarrow)}{D(z,k_\downarrow)}\right)\left(\frac{D(z,k_\downarrow)}{D(z,k)}\right)\delta_m(z,\vk) \nonumber \\
		& = b_1^L(z,M)\mathcal{R}(z,k)\,\delta_m(z,\vk)\;,
	\end{align}
	where $k_\downarrow\gg k_{\pm}$ is a sub-Jeans mode (in the notation of \cite{chiang/etal:2016}), and we have used the scale-independence of the growth rate at $z_i\gg z_\text{DE}$, i.e. 
	$D(z_i,k)\approx D(z_i,k_\downarrow)$. 
	In the last equality, we have rescaled $b_1^L(z_i,M)$ to define
	\begin{equation}
		b_1^L(z,M) = \frac{D(z_i,k_\downarrow)}{D(z,k_\downarrow)} b_1^L(z_i,M) 
	\end{equation} 
	relative to the matter density field linearly extrapolated to redshift $z$ (see e.g. \cite{desjacques/etal:2010}), and we have introduced
	\begin{equation}
		\mathcal{R}(z,k)\equiv \frac{D(z,k_\downarrow)}{D(z,k)} 
	\end{equation}
	to encapsulate the scale-dependence of the halo bias arising from DDE. 
	Therefore, since the evolved DM halo density field is $\delta_h(z,\vk) = \delta_h^L(z_i,\vk) + \delta_m(z,\vk)$ \cite{nusser/davis:1994,fry:1996,mo/white:1996}, we can write 
	$\delta_h(z,\vk) = b_1(z,M,k) \delta_m(z,\vk)$ with
	\begin{equation}
		\label{eq:biaskll}
		b_1(z,M,k) = 1 + b_1^L(z,M)\,\mathcal{R}(z,k) \;.
	\end{equation}
	Eq.~(\ref{eq:biaskll}) is our second estimate for the scale-dependent bias induced by DDE.
	It agrees with the result of \cite{parfrey/hui:2008,parfrey/etal:2011} in the limit of a DM halo formation redshift $z_\text{form}\to\infty$ since the Lagrangian bias approach 
	considered here does not allow for "passive" evolution, by definition.
	On scales $k_\uparrow\ll k_{\pm}$ larger than the Jeans scales of the DE fluids, the linear bias $b_1(z,M,k_\uparrow)$  differs from $b_1(z,M,k_\downarrow)$ by a term $D(z,k_\downarrow)/D(z,k_\uparrow)$. 
	This suppresses $b_1(z,M,k_\uparrow)$ on super-Jeans scale relative to $b_1(z,M,k_\downarrow)$ on sub-Jeans scale if $D(z,k_\downarrow)<D(z,k_\uparrow)$. 
	
	\begin{figure*}[t!]
		\centering
		\includegraphics[width=\textwidth]{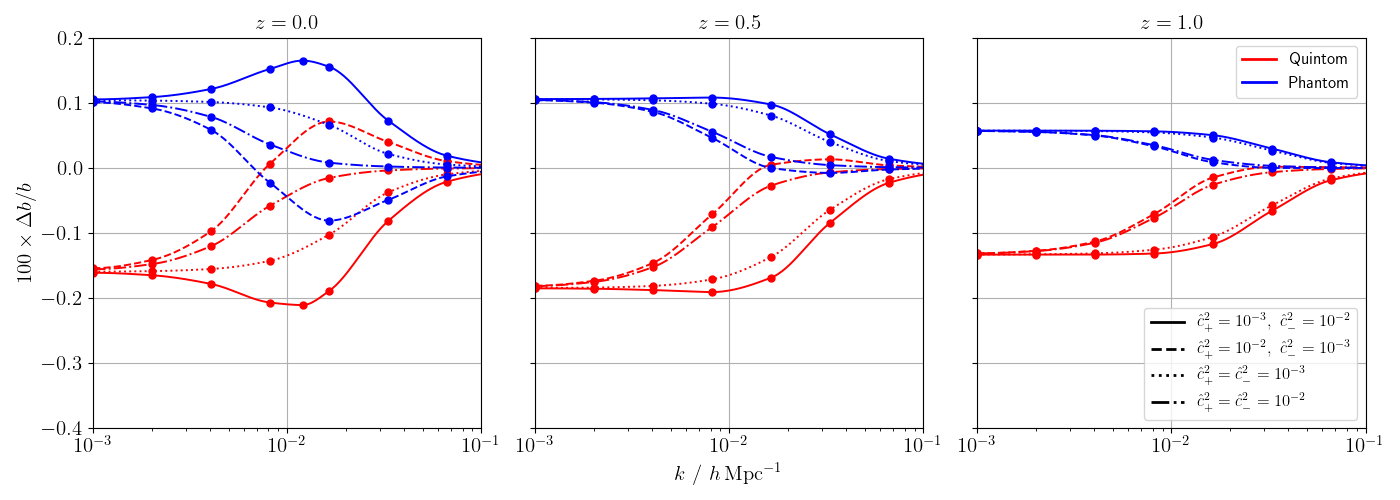}
		\caption{Scale-dependent bias (normalized to unity at $k=k_\downarrow$, here taken to equal $k_\downarrow = 3 \,\hmmpc $) in the 2-fluid Quintom (red) and Phantom (blue) models 
			for a halo mass $M=10^{14}\,\hmsun$, computed with the SU approach. The panels show results at redshift $z=0$ (left), $z=0.5$ (center), and $z=1$ (right).}
		\label{fig:biask}
	\end{figure*}
	
	\subsubsection{Theoretical predictions}
	
	In Fig.~\ref{fig:biask_SUvsLL}, we plot the fractional deviation
	\begin{equation}
		\frac{\Delta b}{b} = \frac{b_1(z,M,k)}{1 + b_1^L(z,M)} - 1
	\end{equation}
	from a scale-independent bias predicted by the SU approach (filled circles with a thin interpolating line) 
	and the local Lagrangian approximation (thick curves) for two of the models shown in Fig.~\ref{fig:growth}.
	We assume a halo mass $M=10^{14}\hmsun$ (corresponding to rich groups hosting a bright central galaxy).
	For this halo mass, the magnitude of the scale-dependence is comparable to that induced by massless neutrinos in the $\Lambda$CDM model 
	\cite{munoz/dvorkin:2018} (see also \cite{LoVerde:2016ahu,Rogozenski:2023tse} for massive neutrinos, which lead to enhanced scale-dependence).
	The local Lagrangian bias approximation overestimates the SU prediction by $\sim 50$\%, in agreement with the findings of \cite{chiang/etal:2016} (see their Fig.11).
	
	Fig.~\ref{fig:biask} illustrates the redshift dependence of $\Delta b /b$, which is shown at redshift $z=0$, 0.5, and 1 for all the models considered in Fig.~\ref{fig:growth}.
	A halo mass $M=10^{14}\hmsun$ is again assumed.
	The scale-dependent bias depends on the value of the DE sound speeds mainly through the change in the Jeans wavenumbers $k_\pm$.
	Lowering the sound speeds somewhat increases the difference between $\Delta b/b$ evaluated at $k=k_\downarrow$ and $k=k_\uparrow$, as well as the steepness of the transition that takes place around $k= k_\pm$. Furthermore, the scale dependence is largest at redshift close to the epoch of phantom crossing ($z \sim 0.2$).
	
	The dominance of one fluid over the other in sourcing scale-dependent dark-energy clustering is determined by their relative contribution $\Omega_-(z)\,\delta_-(k,z)$ and $\Omega_+(z)\,\delta_+(k,z)$ to the total DE energy perturbation. Their ratio depends on the background evolution and the importance of pressure gradients, since a fluid contributes to the perturbation only above its Jeans scale. 
	As shown in Fig.~\ref{fig:omegaevol}, the $w_+$ fluid dominates $\rho_\text{DE}$ throughout most of the evolution, which explains the behavior seen in Fig.~\ref{fig:biask}. The scale-dependent contribution from the $w_-$ fluid becomes relevant only at late time $z\simeq 0$ when its background density increases significantly. 
	This behavior is largely insensitive to the choice of sound speeds. 
	
	If the two EOS parameters are sufficiently close, pressure effects may compensate for differences in the background densities, so that the dominant contribution to perturbations depends on scale. 
	As an example, take a model with $k_+<k_-$ and consider wavenumbers $k_\ast$ satisfying $k_+<k_\ast<k_-$. For such Fourier modes, pressure suppresses perturbations in the $w_+$ fluid but not in the $w_-$ fluid. 
	The corresponding density contrasts scale as $\delta_-(k_\ast,z)\sim(1+w_-)\delta_m$ and $\delta_+(k_\ast,z)\sim(1+w_+)\mathcal{H}^2/(k_\ast^2\hat c_+^2)\,\delta_m$. 
	Therefore, when the EOS parameters are comparable, the $w_-$ fluid dominates the perturbations on these scales. 
	
	\subsection{Extension to higher-order bias parameters}
	
	These approximations can be readily extended to higher-order bias parameters. 
	In the SU approach, for instance, the quadratic Lagrangian bias parameter $b_2^L(k)$ is given by
	\begin{align}
		b_2^L(z,M,k) &= \frac{1}{\bar n}\frac{{\rm d}^2\bar n}{{\rm d}\delta_m^2} \\
		&= - b_1^L(z,M) \frac{{\rm d}^2\delta_c}{{\rm d}\delta_m^2} + b_2^L(z,M) \left(\dv{\delta_c}{\delta_m}\right)^2 \nonumber 
	\end{align}     
	where $b_2^L(z,M)$ is the scale-independent piece of the second-order bias. 
	Solving for the spherical collapse gives $\big\lvert {\rm d}^2\delta_c/{\rm d}\delta_m^2\big\lvert\lesssim 10^{-7}$, which is a few orders of magnitude smaller than the scale-dependent effect in $({\rm d}\delta_c/{\rm d}\delta_m)^2$.
	Therefore, we can set $b_2^L(z,M,k)\approx b_2^L(z,M) ({\rm d}\delta_c/{\rm d}\delta_m)^2$ in a very good approximation. 
	
	\subsection{Detectability}
	
	Since forthcoming surveys will dramatically improve measurements of galaxy clustering statistics, it is interesting to assess the detectability of the scale-dependence induced by the
	DDE sound mode in the power spectrum and bispectrum of galaxy number counts. 
	
	\subsubsection{Single tracer}
	
	We consider first the detectability of the scale-dependence in measurements of the power spectrum of a single galaxy sample. 
	We assume linear theory, ignore redshift-space distortions and consider the model (we omit the redshift dependence hereafter to avoid clutter)
	\begin{equation}
		P_g(k) \equiv b_1^2(k) D^2(k) P_L(k) + \frac{1}{\bar n_g}
	\end{equation}
	where $P_L(k)$ is a reference ($\Lambda$CDM) linear matter power spectrum, $\bar n_g$ is the galaxy number density, $D(k)$ is the scale-dependent linear growth rate and  
	$b_1(k)$ is the linear, scale-dependent galaxy bias computed in the SU approximation following Eq.~(\ref{eq:biaskSU}). 
	Let $D_\downarrow=D(k_\downarrow)$ and $b_{1\downarrow}\equiv b_1(k_\downarrow)$ be the scale-independent piece of the growth rate and linear bias, respectively.
	The difference between $P_g(k)$ and the baseline power spectrum $P_b(k) = b_{1\downarrow}^2 D_\downarrow^2 P_L(k) + \frac{1}{\bar n_g}$,
	\begin{equation}
		\Delta P(k) = P_g(k) - P_b(k) \;,
	\end{equation}
	encodes the scale-dependent signal induced by DDE. 
	
	The signal-to-noise-ratio (SNR) per Fourier mode thus is $\Delta P(k)/\sigma_P(k)$, 
	where the variance $\sigma_P^2(k)=P_g^2(k)$ of the measured power spectrum for a single Fourier mode includes both cosmic variance and shot noise. 
	Summing over all the independent Fourier modes within the survey volume $V_s$ yields
	\begin{equation}
		\label{eq:snr}
		\left(\frac{S}{N}\right)^2 = \frac{V_s}{4\pi^2} \int_{k_{\rm min}}^{k_{\rm max}} {\rm d}k\, k^2 \, \left[ \frac{\Delta P(k)}{\sigma_P(k)} \right]^2.
	\end{equation}
	The lower limit $k_{\rm min}$ corresponds to the largest scales accessible to the survey, while $k_{\rm max}$ denotes the maximum wavenumber beyond which the linear approximation 
	to $P_g(k)$ ceases to be valid. Here we consider a survey with volume $V_s= 10\hhhgpc$ at median redshift $z=0.5$, 
	and set $k_{\rm min}\approx \pi/V_s^{1/3} = 2\cdot 10^{-3}\hmmpc$ and $k_{\rm max}=0.1\hmmpc$. 
	The scale-dependence of $D(k)$ and $b_1(k)$ occurs in the wavenumber range $(k_{\rm min},k_{\rm max})$ if the sound speeds are in the range $10^{-4}\lesssim \hat c_\pm^2\lesssim 10^{-2}$.
	When the errors are cosmic variance dominated, the SNR is largest when the scale-dependence occurs close to $k_\text{max}$ owing to the $k^3$ growth of the number of independent modes. 
	However, for a single galaxy sample with host halo masses in the range $M\sim 10^{12} - 10^{13}\hmsun$ characteristic of luminous red galaxies (LRGs), we find SNR$<0.1$ 
	for all possible values of the sound speeds under consideration. Therefore, detecting this scale dependence is unrealistic unless a multi-tracer analysis is performed.
	We explore this approach in the remainder of this Section.
	
	\subsubsection{Multiple tracers}
	
	By comparing differently biased tracers of the same surveyed volume, multi-tracer analyses can partly alleviate the limitation brought by cosmic 
	variance and improve the detection level of scale-dependent features 
	\cite{seljak:2009,mcdonald/seljak:2009,hamaus/etal:2010,hamaus/etal:2011,hamaus/etal:2012,blake/etal:2013,abramo/leonard:2013,yamauchi/etal:2014}.
	Furthermore, the scale-dependence induced by the sound mode is also present in the bispectrum owing to its dependence on the galaxy bias parameters.
	This motivates a Fisher forecast for a two-tracer measurement of the power spectrum and bispectrum. 
	For simplicity, we will assume that the tracers are central galaxies residing in distinct host DM halos and consider threshold samples characterized by a minimum halo mass $M_\text{min}$.
	
	Following \cite{yamauchi/etal:2017,karagiannis/etal:2024}, the data vector includes the combinations of power spectra and bispectra of tracers "{\it a}" and "{\it b}", i.e.
	\begin{equation}
		\label{eq:datavector}
		\vvx = \{P_{aa},P_{ab},P_{bb},B_{aaa},B_{aab},B_{abb},B_{bbb} \}
	\end{equation}
	where $P_{ij}=P_{ij}(k)$ and $B_{ijk}=B_{ijk}(k_1,k_2,k_3)$, with $(k_1,k_2,k_3)$ satisfying the triangle condition
	and $B_{ijk}$ symmetrized in order to avoid double counting of information \cite{karagiannis/etal:2024}.
	In the tree-level, real-space approximation considered here, the power spectra read 
	\begin{equation}
		P_{ij}(k) = b_1^{(i)}\!(k)\, b_1^{(j)}\!(k)\, D^2(k)\, P_L(k) + \frac{\delta^{\rm K}_{ij}}{\bar n_i} \;,
	\end{equation}
	where $b_1^{(i)}(k)$ and $\bar n_i$ are the linear scale-dependent bias and the number density of the galaxy sample "{\it i}".
	Likewise, the auto and (symmetrized) cross-bispectrum $B_{aaa}(k_1,k_2,k_3)$ and $B_{aab}(k_1,k_2,k_3)$ are given by
	\begin{align}
		B_{aaa} &= \Big\{2\,Z_1^a(k_1)\,Z_1^a(k_2)\, Z_2^a(\vk_1,\vk_2)\,D^2(k_1) D^2(k_2) \\
		&\qquad \times P_L(k_1)\,P_L(k_2) + \mbox{(2 cyc.)} \Big\} \nonumber \\
		&\quad + 2\, P_{\epsilon\epsilon_\delta}^a\Big[Z_1^a(k_1)D^2(k_1) P_L(k_1) +\mbox{(2 cyc.)}\Big] + B_{\epsilon\epsilon\epsilon}^a \nonumber 
	\end{align}
	and 
	\begin{align}
		B_{aab} &= \frac{2}{3}\Big\{ \Big[\big(Z_1^a(k_1) Z_1^b(k_2) + Z_1^b(k_1) Z_1^a(k_2)\big) Z_2^a(\vk_1,\vk_2) \nonumber \\
		&\qquad + Z_1^a(k_1) Z_1^a(k_2) Z_2^b(\vk_1,\vk_2)\Big] \nonumber \\
		&\quad \times D^2(k_1) D^2(k_2)\, P_L(k_1)P_L(k_2) + \mbox{(2 cyc.)} \Big\} \nonumber \\
		&\quad + \frac{1}{3}P_{\epsilon\epsilon_\delta}^a\Big[Z_1^b(k_1)D^2(k_1) P_L(k_1)+\mbox{(2 cyc.)}\Big] \;,
	\end{align}
	while $B_{abb}$ and $B_{bbb}$ follow from the replacement $a\leftrightarrow b$. 
	Here, "(2 cyc.)" denotes the cyclic permutations of $k_\alpha\equiv (k_1,k_2,k_3)$. 
	
	We approximate the stochastic contributions to the bispectrum by the Poisson expectations \cite{pollack/smith/porciani:2012,Schmidt:2015gwz,chan/blot:2017}
	(i.e. we ignore the halo exclusion effects discussed in \cite{hamaus/etal:2010,ginzburg/etal:2017})
	\begin{equation}
		P_{\epsilon\epsilon_\delta}^i = \frac{b_{1\downarrow}^{(i)}}{2 \bar n_i}\;,\quad B_{\epsilon\epsilon\epsilon}^i = \frac{1}{\bar n_i^2} \;,
	\end{equation}
	where again $b_{1\downarrow}^{(i)}$ is the scale-independent contribution to the linear bias of sample "{\it i}".
	In the absence of redshift-space distortions, the kernels $Z_1^i$ and $Z_2^i$ simplify to $Z_1^i(\vk)= b_1^{(i)}\!(k)$ and 
	\begin{align}
		Z_2^i(\vk_1,\vk_2) &= b_1^{(i)}\!(k_3)\, F_2(\vk_1,\vk_2) + \frac{1}{2}\, b_2^{(i)}\!(k_3) \nonumber \\
		&\qquad + b_{K^2}^{(i)}\!(k_3)\, S_2(\vk_1,\vk_2) \;.
	\end{align}
	Explicit expressions for $F_2(\vk_1,\vk_2)$ and $S_2(\vk_1,\vk_2)$ can be found in \cite{ptreview,biasreview}. 
	The explicit argument structure of $Z_2$, $F_2$, and $S_2$ emphasizes that these kernels depend on the full triangle configuration $k_\alpha$ in a non-symmetric way. 
	The bias parameters $b_2^{(i)}\!(k)$ and $b_{K^2}^{(i)}\!(k)$ associated with the dependence of the galaxy number counts on $\delta_m^2$ and on the squared (traceless) 
	tidal shear $K^2=K_{ij}K^{ij}$ (where $K_{ij}=\partial_i\partial_j\Phi-(1/3)\delta_{ij}\nabla^2\Phi$) becomes scale-dependent in the presence of the DE sound mode.
	To calculate their scale-dependence, we assume a one-to-one correspondence between the scale-independent bias parameters $b_{1\downarrow}^{(i)}$ and $b_{2\downarrow}^{(i)}$ 
	given by~\cite{lazeyras/etal:2016} 
	\begin{equation}
		\label{eq:biasrelation1}
		b_{2\downarrow}^{(i)} = 0.412 -  2.143\, b_{1\downarrow}^{(i)} + 0.929 \big(b_{1\downarrow}^{(i)}\big)^2 + 0.008 \big(b_{1\downarrow}^{(i)}\big)^3 \;,
	\end{equation}
	and a relation between $b_{K^2}^{(i)}\!(k)$ and $b_1^{(i)}\!(k)$ arising from co-evolution~\cite{catelan/etal:1998,catelan/etal:2000,chan/scoccimarro/sheth:2012,baldauf/etal:2012},
	\begin{equation}
		\label{eq:biasrelation2}
		b_{K^2}^{(i)}\!(k) = -\frac{2}{7}\big(b_1^{(i)}\!(k) - 1\big) \;.
	\end{equation}
	Note, however, that deviations from co-evolution are expected theoretically and measured in numerical simulations \cite{sheth/etal:2013,modi/etal:2017}. 
	These assumptions lead to
	\begin{align}
		b_2^{(i)}\!(k) &= -\frac{8}{21} \left(b_{1\downarrow}^{(i)} -1\right) \dv{\delta_c}{\delta_m} \\
		&\quad  +\left(b_{2\downarrow}^{(i)}-\frac{8}{21}\big(b_{1\downarrow}^{(i)}-1\big)\right)\left(\dv{\delta_c}{\delta_m} \right)^2 
		\nonumber \\
		b_{K^2}^{(i)}\!(k) &= \frac{2}{7}\left(b_{1\downarrow}^{(i)}-1\right) \dv{\delta_c}{\delta_m} \nonumber \;,
	\end{align}
	which depend only on $b_{1\downarrow}^{(i)}$ and on ${\rm d}\delta_c/{\rm d}\delta_m$. 
	Note that the Eulerian-to-Lagrangian relation $b_2 = \frac{8}{21} b_1^L + b_2^L$ was used to derive $b_2^{(i)}\!(k)$.
	
	\begin{figure}
		\centering
		\includegraphics[width=.48\textwidth]{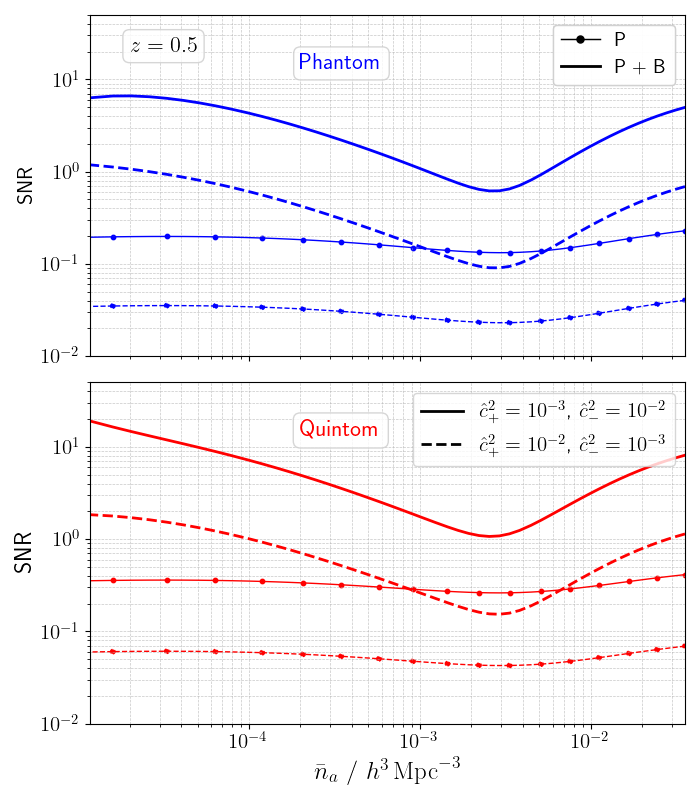}
		\caption{Signal-to-noise-ratio (SNR) for a detection of the scale dependence induced by the DDE sound mode in a power spectrum (P) and power spectrum + 
			bispectrum (P+B) analysis of two different tracers. The SNR is plotted as a function of the number density $\bar n_a$ of tracer "{\it a}", with 
			$b_{1\downarrow}^{(a)}$ varied consistently with $\bar n_a$ (see text for details). 
			Results are shown for the Phantom (top) and the Quintom (bottom) models assuming a survey volume $V_s=10\hhhgpc$ and two different choices of sound speed $\hat c_\pm$, see footnote ~\ref{footnote}.}
		\label{fig:snr_phantom}
	\end{figure}
	
	Let $\vvx_P=\{P_{aa},P_{ab},P_{bb}\}$ be the data vector of the power spectra measured as a function of $k$.
	The Fisher matrix derived from $\vvx_P$ for a set of parameters $\{p_i\}$ is \cite{vogeley/szalay:1996,tegmark/etal:1998}
	\begin{equation}
		\label{eq:FisherP}
		F_{ij}^P= \frac{V_s}{2\pi^2}\int\!\!{\rm d}k\,k^2\,\frac{\partial\vvx_P(k)}{\partial p_i} {\rm Cov}_P^{-1}(k) \frac{\partial\vvx_P^\top(k)}{\partial p_j} \;,
	\end{equation}
	where the covariance matrix ${\rm Cov}_P(k)$ of $\vvx_P(k)$ is given in Appendix \S\ref{app:covariance}.
	Although the covariance of the full data vector $\vvx$ (see Eq.~\ref{eq:datavector}) includes the covariance between power spectra and bispectra measurements, 
	we will neglect this cross-covariance for simplicity (see \cite{song/etal:2015,yankelevich/porciani:2019,karagiannis/etal:2024} for the impact of the cross-covariance).
	Consequently, the Fisher matrix derived from $\vvx$ reduces to $F_{ij}^{P+B}=F_{ij}^P + F_{ij}^B$, 
	with \cite{yamauchi/etal:2017,yankelevich/porciani:2019,karagiannis/etal:2024}
	\begin{align}
		\label{eq:FisherB}
		F_{ij}^B &= \frac{3 V_s}{8\pi^4}\int_{\Delta}\!\! {\rm d}k_1 {\rm d}k_2 {\rm d}k_3\, k_1 k_2 k_3\,s_{123}^{-1} \\
		&\qquad \times \frac{\partial\vvx_B(k_\alpha)}{\partial p_i} {\rm Cov}_B^{-1}(k_\alpha) \frac{\partial\vvx_B^\top(k_\alpha)}{\partial p_j} \nonumber
	\end{align}
	in the continuum limit.
	The domain of integration $\Delta=\Delta(k_1,k_2,k_3)$ is defined by the ordering $k_1\leq k_2\leq k_3$ and the triangle 
	condition $|k_1-k_2|\leq k_3\leq k_1+k_2$ \cite{scoccimarro/etal:1998,sefusatti/etal:2006}, while $s_{123}$ is the usual triangle symmetry factor
	($s_{123}=1$, 2 and 6 for scalene, isosceles and equilateral triangles, respectively) \cite{fry/etal:1993,scoccimarro/etal:2004}.
	Alternatively, one could integrate all $k_i$ over the full range $[k_\text{min},k_{\rm max}]$ and replace $s_{123}$ by $3!$.
	The vector $\vvx_B=\{B_{aaa},B_{aab},B_{abb},B_{bbb}\}$ of bispectrum measurements is a function of the full triangle configuration $k_\alpha$.
	The corresponding covariance matrix ${\rm Cov}_B(k_\alpha)$ is given in Appendix \S\ref{app:covariance}.
	
	To calculate the SNR for a detection of the scale dependence induced by the DDE sound mode, we fix the shape and amplitude of the reference power spectrum
	$P_L(k)$, assume the bias relations (\ref{eq:biasrelation1}) and (\ref{eq:biasrelation2}), 
	and consider the three-parameter set ${\bf p}=\{b_{1\downarrow}^{(a)},b_{1\downarrow}^{(b)},A\}$. 
	Here, $b_{1\downarrow}^{(a)}$, $b_{1\downarrow}^{(b)}$ are the (scale-independent) linear bias of the two tracers "{\it a}" and "{\it b}",
	while $A$ parameterizes the scale dependence in the growth rate $D(k)$ and the response $({\rm d}\delta_c/{\rm d}\delta_m)(k)$ according to
	\begin{align}
		D(k) &= D_\downarrow \Big(1+ A\, f_1(k)\Big) \\
		\dv{\delta_c}{\delta_m}(k) &= -1 + A\, f_2(k) \nonumber \;.
	\end{align}
	Note that the auxiliary functions $f_1(k)$ and $f_2(k)$ are derived separately from the linear growth equation and from the SU approach, respectively.
	Since the actual signal has $A\equiv 1$,
	the SNR for a detection of the sound mode is thus defined as SNR=$\sigma_A^{-1}$ where $\sigma_A$ is the marginalized uncertainty on $A$. 
	Furthermore, we vary $b_{1\downarrow}^{(a)}$ while keeping $b_{1\downarrow}^{(b)}=1$ fixed to forecast detection limits as a function of the tracers properties.
	The number density $\bar n_a$ and $\bar n_b$ (of central galaxies) are obtained by matching them to the cumulative halo number density evaluated at the minimum 
	halo mass $M_\text{min}$ that reproduces the desired linear bias values $b_{1\downarrow}^{(a)}$ and $b_{1\downarrow}^{(b)}$.
	Therefore, the larger $b_{1\downarrow}^{(i)}$ the smaller $\bar n_i$ as expected from a peak-background split argument \cite{kaiser:1984,bbks,mo/white:1996,sheth/tormen:1999}.
	At redshift $z=0.5$, this procedure yields $\bar n_b\simeq 0.01\hhhmpc$ for $b_{1\downarrow}^{(b)}=1$, adopting the halo mass function and bias fits from \cite{Tinker10}.
	These are the number density and linear bias that we will hereafter adopt for sample "{\it b}".
	The value of $b_{2\downarrow}^{(i)}$ is determined by Eq.~(\ref{eq:biasrelation1})~\footnote{For the threshold samples $M\geq M_\text{min}$ considered here, the bias 
		relation should, in principle, be integrated over the halo mass $M$ to infer $b_{2\downarrow}^{(i)}$. We have found that this procedure increases $b_{2\downarrow}^{(i)}$
		by $\sim$20\%, thereby enhancing the scale-dependence of the bispectrum. Hence our SNR predictions are conservative here.}
	
	Fig.~\ref{fig:snr_phantom} shows the SNR for a detection of the scale dependence as a function of the number density $\bar n_a$ for both the Phantom and the Quintom models
	with two different choices of sound speed as indicated in the figure~\footnote{\label{footnote}The SNR forecasted for the model with $\hat c_\pm^2=10^{-2}$ (resp. $\hat c_\pm^2=10^{-3}$) is very similar to the SNR shown in Fig.~\ref{fig:snr_phantom} 
		for $(\hat c_+^2,\hat c_-^2)=(10^{-2},10^{-3})$ (resp. $(\hat c_+^2,\hat c_-^2)=(10^{-3},10^{-2})$)}.
	Results are shown for a power spectrum only (P) and a joint power spectrum and bispectrum (P+B) analysis assuming $V_s=10\hhhgpc$. The plotted range of $\bar n_a$ corresponds to a range in minimal mass $M_\text{min}$ of $10^{11} - 10^{14} \hmsun$, from right to left. 
	Adding bispectrum measurements is essential to achieve an SNR larger than unity. 
	In the P+B analysis, the SNR reaches a minimum at the number density $n_a \sim 2\times 10^{-3}\hhhmpc$ corresponding to an unbiased sample $b_{1\downarrow}^{(a)}\sim 1$
	(i.e. no scale-dependent bias).
	The rise in SNR toward low and high values of $\bar n_a$ is driven by the increase of $|b_{1\downarrow}^{(a)} - 1|$, which amplifies the scale dependence of the bispectrum 
	mainly through the bias functions $b_1^{(a)}\!(k)$ and $b_2^{(a)}\!(k)$. 
	For a sample density $\bar n_a\approx 10^{-4}\hhhmpc$ for instance, which yields $b_{1\downarrow}^{(a)}\approx 3.5$, the scale dependence induced by the DDE sound mode 
	is detected with a ${\rm SNR}\sim 4$ assuming $(\hat c_+^2,\hat c_-^2)=(10^{-3},10^{-2})$ in the Phantom model. 
	Overall, the SNR is predominantly sensitive to the value of $\hat c_+^2$ (since the $w_+$ fluid
	dominates the DE energy density at redshift $z\sim 0.5$). As $\hat c_+^2$ decreases from $10^{-2}$ to $10^{-3}$, the characteristic scale dependence shifts to larger $k$,
	thereby increasing the number of accessible wavemodes and, consequently, the SNR. 
	In the Quintom model, the SNR is generally higher owing to the larger scale dependence of $D(k)$ and $b_1^{(i)}\!(k)$ (see Figs .~\ref{fig:growth} and \ref{fig:biask}). 
	Note that the decrease of the SNR in the Phantom model for $c_+^2 = 10^{-3}$ at very low $n_a$ may be an artifact of the bias relation Eq.~\eqref{eq:biasrelation1}.
	
	\begin{figure}
		\centering
		\includegraphics[width=.48\textwidth]{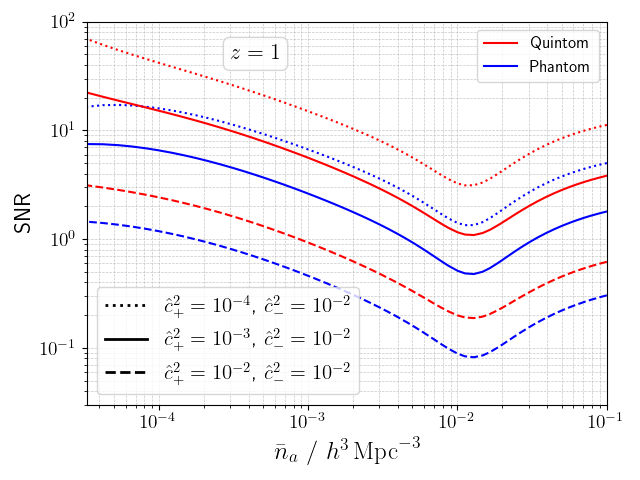}
		\caption{SNR for the detection of the scale dependence induced by the DDE sound mode from a P+B analysis of two tracers at redshift $z=1$, assuming a survey volume 
			$V_s=20\hhhgpc$.}
		\label{fig:snr_z1}
	\end{figure}
	
	Fig.~\ref{fig:snr_z1} shows the SNR from a P+B analysis of two tracers at $z=1$ assuming a survey volume $V_s=20\hhhgpc$. 
	Theoretical curves are shown for $10^{-4}\leq \hat c_+^2\leq 10^{-2}$ and a fixed sound speed $\hat c_-^2=10^{-2}$ (which has a very small impact on the results since 
	the $w_-$ fluid is completely negligible at $z=1$).
	Here again, $\bar n_a$ and $b_{1\downarrow}^{(a)}$ are varied consistently using the fitting formulae from \cite{Tinker10} under the assumption that each halo host exactly one
	surveyed (central) galaxy, whereas the second sample "{\it b}" has number density $\bar n_b=0.01\hhhmpc$ and bias $b_{1\downarrow}^{(b)}=1$ as in Fig. ~\ref{fig:snr_phantom}.  In this case the plotted range of $\bar n_a$ corresponds to a range in minimal mass $M_\text{min}$ of $10^{10.5} - 10^{13.5} \hmsun$, from right to left. 
	The maximum wavenumber is $k_\text{max}=0.1\hmmpc$, which captures most of the scale-dependence in the limiting case of $c_+^2 = 10^{-4}$. For lower sound speeds, one should resort to non-linear extensions in order to probe higher $k$-values. 
	At fixed $\bar n_a$, the weakening of the scale dependence of $D(k)$ and $b_1^{(a)}(k)$ with increasing redshift is compensated by larger values of $b_{1\downarrow}^{(a)}$.
	However, detecting sound speeds $\hat c_+^2>10^{-2}$ at $z=1$ still requires a highly biased sample "{\it a}" with number density $\bar n_a\lesssim 10^{-4}\hhhmpc$ and bias 
	$b_{1\downarrow}^{(a)}\gtrsim 5$. 
	
	To conclude, we note that the rise in the SNR at small $\bar n_a$, i.e. large $b_{1\downarrow}^{(a)}$, suggests that weighting the tracers by halo mass as advocated in
	\cite{seljak/etal:2009,hamaus/etal:2010,hamaus/etal:2012} should improve further the detectability of the scale dependence induced by the DDE sound mode.

	\section{Gravitational drag}
	\label{sec:df}
	
	In addition to the scale-dependent growth, DDE also acts as a dissipative medium for the DM halos and thus affects their velocities. 
	In analogy with the baryon-CDM perturbations \cite{blazek/etal:2016,schmidt:2016}, DDE generates a linear-order contribution to the halo velocities 
	of the form
	\be
	\vv_h(z,\vx) = \vv_m(z,\vx) + \beta_v^{\text{DE}}(z)\,\vv_r(z,\vx) + \dots 
	\label{eq:vh}
	\ee
	where $\beta_v^\text{DE}(z)$ is a velocity bias parameter and $\vv_r(z,\vx)=\vv_{m}(z,\vx)-\vv_\text{DE}(z,\vx)$ is the relative velocity between the
	matter and DE distribution (see \cite{biasreview} for a review).
	The physical origin of this contribution is the dynamical friction (DF) exerted by DDE on the DM halos as they move relative to the DDE frame. 
	This relative motion excites sound modes and generates a density wake in the DDE component, which can slow their motion. 
	
	\subsection{Magnitude estimate}
	
	A rough estimate of the magnitude of this gravitational drag can be obtained as follows. 
	The magnitude of the DF force $F_\pm$ produced by either component of the DDE is given by
	\begin{equation}
		F_\pm= 4\pi \frac{(GM)^2}{v_r^2}\rho_\pm I_\pm(v_r)\;,
	\end{equation}
	where $M$ is the DM halo mass, $v_r$ is the velocity of the DM halo relative to the DE fluids, 
	and the friction coefficient satisfies $I_\pm(v_r)\approx (1/3) (v_r/\hat c_\pm)^3$ in the small-velocity limit $v_r\ll\hat c_\pm$ relevant here 
	(note that there is no Coulomb logarithm in this regime). 
	The energy dissipated by DF at redshift $z$ is $\Delta E=F_\pm v_h t_H$, where $v_h$ is the velocity of the DM halo (in e.g. the CMB frame). This leads to
	\begin{align}
		\frac{\Delta E}{E} &\simeq 0.5 \times \left(\frac{M}{10^{15}\hmsun}\right)\left(\frac{10^{-5}}{\hat c_\pm^2}\right)^{3/2}\left(\frac{v_r}{v_h}\right) \\
		&\qquad \times \mathcal{E}(z) \,\Omega_\pm(z)\, \Delta_\pm \nonumber \;.
	\end{align}
	Here, $E=(1/2) M v_h^2$ is the kinetic energy of the DM halo, $\mathcal{E}(z)=H(z)/H_0$ and $\Delta_\pm = \delta_\pm(k_\pm,z)$ is the fractional density perturbation of either of 
	the two DE fluids in a patch of size $\sim k_{\pm}^{-1}$ in which the DM halo is moving.
	In the redshift range $z\lesssim 2$ we have $\Omega_\Lambda(z)\mathcal{E}(z)\sim \mathcal{O}(1)$. Therefore, the effect is negligible unless (at least) one of the squared sound 
	speeds approaches $\hat{c}_\pm^2 \sim 10^{-5}$. 
	If this happens, the drag modifies the DM halo velocities and, thereby, measurements of the growth rate $f\sigma_8$ inferred from redshift space distortions (RSD), 
	with the size of the effect largely determined by the halo mass and the sound speeds of the DE fluids.
	Let us illustrate this point below. 
	
	\subsection{Zel'dovich approximation}
	
	We model the growing DM halo as the extended (proper) density perturbation
	\begin{equation}
		\rho_h(z,\vx) = \frac{1}{a^3(z)} M(z)\,u_h\big(|\vx - \vx_h(z)|,z\big)\;,
	\end{equation}
	where $\vx_h(z)$ is the comoving position of the DM halo, $M(z)$ is the time-dependent halo mass and $u_h(r,z)$ is a spherically symmetric density profile normalized 
	such that $\int\!\dd^3\vr\,u_h(r,z) = 1$. Since finite-size effects are negligible in the subsonic regime $v_h/\hat c_\pm\ll 1$ considered here 
	(see Appendix \ref{app:finite_size} for a justification), we will treat the DM halos as point particles in what follows.
	In linear theory, the comoving position $\vx_h(z)$ and peculiar velocity $\vv_h(z)$ of DM halos follow the Zel'dovich approximation
	\begin{align}
		\label{eq:Zeldovichmotion}
		\vx_h(z) &= \vq + D(z)\, x_0\, \nvh \\
		\vv_h(z) &= \dot D(z)\, x_0\,\nvh \nonumber \;,
	\end{align}
	where $\vq$ is the Lagrangian ($\eta=0$) halo position, $D(z)$ is the linear growth rate of matter fluctuations (we have ignored the $k$-dependence) 
	and $x_0>0$ is a random variable such that $3x_0^2/\sigma_v^2$ is distributed as $\chi^2$ with 3 degrees of freedom. 
	The 1-dimensional, linear velocity dispersion $\sigma_v$ of DM halos is at the present epoch. 
	The mass growth history $M(\eta)$ is modeled following the approach of \cite{Correa:2014xma} based on the extended Press-Schechter (EPS) formalism 
	(see Appendix \ref{sec:mass_accretion}, as well as \cite{Reed:2003sq,Genel2010,Bosch2014}). 
	
	\begin{figure*}[!t]
		\centering
		\includegraphics[width=1\textwidth]{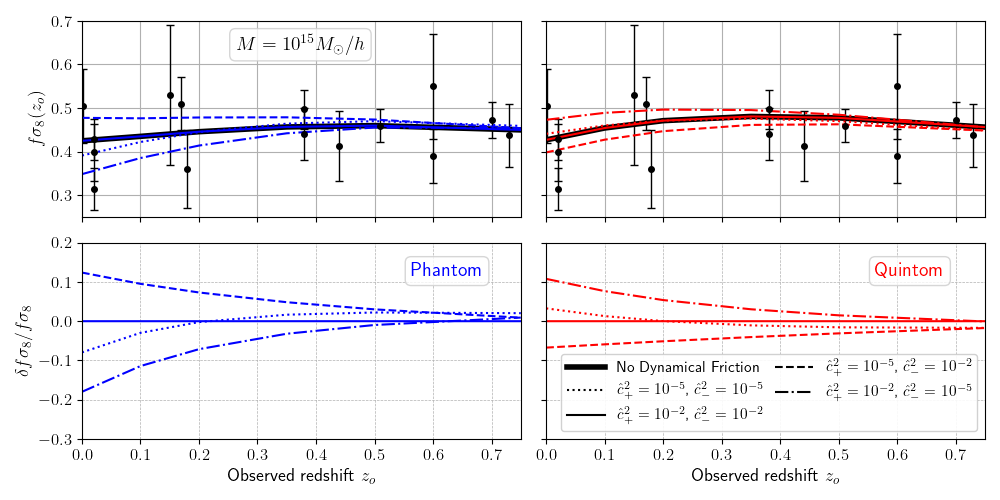}
		\caption{The thin (colored) curves show the normalized growth rate $f_\text{eff}\sigma_8$ that would be inferred from measurements of halo peculiar velocities 
			in a Phantom and Quintom model with (at least one) low sound speed. The thick (black) curve represents the unperturbed Zel'dovich solution (i.e. in the absence 
			of DF).
			Results are shown assuming a halo mass $M = 10^{15} \hmsun$ at the observed redshift $z_o$. 
			Data points indicate actual measurements of $f\sigma_8$ extracted from peculiar velocity and redshift-space distortions (see text for details). 
		}
		\label{fig:m15}
	\end{figure*}
	
	\subsection{Halo velocity evolution}
	
	The gravitational pull of the DM halos sources perturbations in the DE fluids. 
	The size of the resulting sound wake does not exceed the sound horizon of the DE fluids, which is noticeably smaller than the Hubble radius for $\hat c_\pm \ll 1$.
	Since the peculiar velocities of DM halos change on a timescale comparable to the Hubble time $t_H$, the force exerted by the DE fluids on the DM halo at a given time 
	can be approximated by the solution found by \cite{Ostriker:1998fa} for a linearly-moving perturber in an ideal fluid. 
	For subsonic motions, this solution does not depend on the time elapsed since the perturber was turned on. 
	The additional acceleration experienced by the DM halos can thus be approximated by 
	\begin{align}
		\label{eq:accDE}
		{\bf a}_\text{DE}(\eta) &\approx -\frac{(4\pi G)^2}{v^2} M_h(\eta) \Big[\bar\rho_+(\eta) (1+w_+)I_+(v) \\
		&\qquad +\bar\rho_-(\eta)(1+w_-)I_-(v)\Big]\, \nvh\nonumber \\
		&\equiv -a_\text{DE}(\eta,v)\, \nvh \nonumber \;,
	\end{align}
	where the friction coefficients $I_\pm(v)$ behave like $I_\pm(v)\approx (1/3) (v/\hat c_\pm)^3$ in the subsonic regime $v\lesssim \hat c_\pm$
	relevant here. 
	
	For a DM halo moving with an initial velocity $\propto \dot D x_0$, momentum conservation leads to the equation of motion 
	\begin{equation}
		\label{eq:vEoM}
		\dv{\eta} \left( a v_h \right) = a\big(\ddot{D}+aH\dot{D}\big)x_0 - a^2\, a_\text{DE}(\eta,v_h) \;.
	\end{equation}
	The solution $v_h(\eta)$ scales linearly with $x_0$.
	In the limit $a_\text{DE} \to 0$, it reduces to the classical Zel'dovich approximation Eq.~\eqref{eq:Zeldovichmotion}. 
	Deviations from this unperturbed Zel'dovich trajectory scale as 
	\begin{equation}
		\frac{\delta v}{v} \sim 0.1 \times \left(\frac{M}{10^{15}\hmsun}\right) \bigg(\frac{10^{-5}}{\hat c_\pm^2}\bigg)^{3/2}
	\end{equation} 
	at e.g. redshift $z=0.5$.
	
	In the Phantom model, both EOS parameters satisfy $w_\pm<-1$. However, since $\rho_-<0$, the gravitational interaction with the DM halos results in a drag applied by the $w_-$ fluid and a thrust applied by the $w_+$ fluid. By contrast, in the Quintom model, both energy densities are positive ($\rho_\pm>0$). The choice $w_+>-1$ and $w_-<-1$ thus leads to a drag from the $w_+$ fluid and a thrust from the $w_-$ fluid. The overall nature of the gravitational drag depends on the choice of $\hat c_\pm$. To produce a thrust, energy must be extracted from the gravitational interaction between the DM halo and the sound wake. This could occur only if the sound wake generates an underdensity trailing the DM halo (which does not happen for regular matter). 
	
	\subsection{Observed growth rate}
	
	By solving Eq.~\eqref{eq:vEoM} numerically, one can use $v_h(\eta)$ to define an effective growth rate $D_\text{eff}(\eta)$ via the relation
	\begin{equation}
		\dv{\eta} \big[a v_h(\eta)\big] = a(\ddot{D}_\text{eff}+\cH\dot{D}_\text{eff})x_0\;,
	\end{equation}
	which includes the DF experienced by DM halos.
	Since $v_h(\eta)$ scales linearly with $x_0$, it is enough to find $v_h(\eta)$ for a particular choice of $x_0$ so long as the condition $v_h(\eta) \ll \hat c_\pm$ is satisfied 
	throughout the time evolution. The latter holds for most of the parameter space, which justifies the assumption that DM halos move with subsonic velocities. 
	
	The theoretical curves in Fig.~\ref{fig:m15} show the quantity $f_{\rm eff}\sigma_8$
	that would be inferred from (direct or indirect) measurements of DM halo velocities for cluster-size objects of mass $M = 10^{15}\hmsun$ at redshift $z\leq 0.75$. 
	They depend only weakly on the details of the halo mass accretion history (see Appendix \ref{sec:mass_accretion}). 
	Here, both
	\begin{equation}
		f_{\rm eff}(z) \equiv \dv{\ln D_\text{eff}}{\ln a}
	\end{equation}
	and $\sigma_8(z)\propto D_\text{eff}(z)$ depend on the effective growth rate $D_\text{eff}(z)$ and, therefore, are also mass-dependent.
	The data points are a compilation of peculiar velocity and redshift-space distortions measurements of $f\sigma_8$
	\cite{eBOSS:2020yzd,Song:2008qt,Howlett:2017asq,Huterer:2016uyq,Hudson2012,Turnbull2012,Davis:2010sw,Blake:2013nif,Blake2012,Pezzotta:2016gbo,Okumura:2015lvp,eBOSS:2018yfg}. 
	In the extreme cases shown in Fig.~\ref{fig:m15}, the DF induced by the sound mode is significant enough that it affects measurements of $f\sigma_8$ inferred from peculiar
	velocities of rich groups and clusters at a level comparable to the scatter of the measurements. This suggests that $\hat c_\pm\sim 10^{-5}$ is an effective lower bound on the 
	sound speed of the DE fluids. 
	However, since the effect scales as $\delta f/f \sim \delta v/v$, it is extremely hard to probe (much) larger values of $\hat c_\pm^2$ using such measurements.
	
	\subsection{Detectability from stacked weak lensing}
	
	One may ask whether the density wakes produced by the motion of halos through the DE fluids can be detected in stacked weak-lensing measurements. 
	To estimate the number of galaxies needed for a detection, we consider the case in which the $w_+$ fluid dominates the DE sector, so that the induced wake corresponds to a positive density perturbation. 
	In the linear-response regime, the wake generated by a halo of mass $M$ moving through a perfect fluid with sound speed $c_+$ scales as \cite{Eytan:2024yyi}
	\begin{equation}
		\frac{\delta \rho_{\rm wake}(r)}{\bar\rho} \sim \frac{G M}{c_+^2\, r} 
	\end{equation}
	up to $\mathcal{O}(1)$ angular factors. 
	After projection along the line of sight, the characteristic surface mass density inherits the same scaling 
	\begin{equation}
		\Sigma_{\rm wake} \sim \bar\rho \,\frac{G M}{c_+^2}
	\end{equation}
	if one neglects a multiplicative function of the aperture radius $R$ \cite[e.g.][]{Bartelmann:1999yn}. 
	The corresponding weak lensing convergence is
	\begin{equation}
		\kappa \sim \frac{\Sigma_{\rm wake}}{\Sigma_{\rm crit}}\sim \frac{H_0}{c}\,\frac{G M}{c_+^2}
	\end{equation}
	where, in the second equality, we have assumed that the lens and the source are cosmological so that the angular diameter distances are all of order $c/H_0$ and
	the critical surface mass density $\Sigma_{\rm crit}$ is given by
	\begin{equation}
		\Sigma_c \sim \frac{c }{H_0}\bar\rho\;.
	\end{equation}
	For a halo mass $M \sim 10^{13}\hmsun$ approriate for LRGs at redshift $z\sim 0.5$ and a DE sound speed $c_+^2 \sim 10^{-5}$, this gives a characteristic signal 
	of order $\kappa \sim 10^{-5}$ per individual halo. 
	For a conservative intrinsic shape noise (galaxy ellipticity dispersion) $\sigma_e \sim 0.3$, stacked weak lensing measurements would therefore require of order 
	\begin{equation}
		N \sim \left(\frac{\sigma_e}{\kappa}\right)^2 \sim 10^9 
	\end{equation}
	source-lens pairs to achieve a signal-to-noise of order unity, which should be possible with Stage-IV surveys.
	
	\clearpage
	\section{Conclusions}
	\label{sec:conclusions}
	
	In this paper, we have considered a two-fluid dynamical dark energy (DDE) model based on the proposal of \cite{hu:2005}, 
	which allows for a crossing of the phantom divide ($w=-1$) consistent with the results reported by DESI \cite{DESI:2024hhd,DESI:2024mwx,DESI:2025zgx,DESI:2025gwf}.
	Unlike a cosmological constant, DDE can support propagating perturbation modes, which include sound waves in an effective (hydrodynamical) description. 
	If future data confirm the current DESI measurement of the DE equation of state, 
	the detection of a sound mode would provide a smoking gun for its dynamical nature.
	The excitation and observational relevance of this mode depend sensitively on the underlying microphysics, including the EOS parameters $w_\pm$ 
	and effective sound speeds $\hat c_\pm$ of the individual DE fluids. 
	The latter generally is a function of position, or wavenumber, because it includes both adiabatic and non-adiabatic contributions to the pressure perturbation. 
	For simplicity, however, we have assumed constant speeds of sound throughout this work. 
	
	The presence of DDE perturbations introduces a scale dependence in the growth of matter density fluctuations $D(k)$ around the Jeans scales of the DE fluids. 
	This effect arises because perturbations in the DE fluids do not cluster on scales smaller than their sound horizon. 
	This scale-dependent growth leads to a scale-dependent halo bias $b_1(k)$ that can leave observable imprints in the clustering of large-scale structure tracers. 
	We have computed these effects consistently using the linear growth equation and the spherical collapse model. 
	The scale dependence of the halo bias computed in the separate universe approach is approximately 50\% smaller than a prediction based on a local Lagrangian assumption, 
	in agreement with the findings of \cite{chiang/etal:2016}. 
	Although the individual EOS parameters $w_\pm$ significantly differ from $w=-1$ in DDE models considered here, the resulting scale-dependent bias remains small due to subtle 
	cancellations between the two DE components.
	For cluster-size halos ($M\sim 10^{14}\hmsun$), the amplitude of this effect is comparable to that induced by massless neutrinos in the $\Lambda$CDM model.
	Moreover, the sign of the scale dependence in $D(k)$ and $b_1(k)$ depends on the direction in which the phantom divide is crossed. 
	One should expect some degeneracy with the scale dependence induced by neutrinos. However, since this scale dependence is known (up to the uncertainty in the sum 
	of neutrino masses), measurements at different scales and redshifts could help break this degeneracy (at least partially), and isolate the effect arising from DDE. 
	Conversely, the scale dependence induced by DDE would act as a systematic in measurements of the neutrino masses.
	
	To assess the extent to which measurements of galaxy clustering can probe the scale-dependent growth and bias induced by the sound mode of DDE, we have performed a Fisher
	forecast for a combined two-tracer analysis, since a single tracer is not sufficiently sensitive to these effects. 
	Measurements of the auto- and cross-bispectra of the samples are essential to detect the scale dependence induced by the sound mode of DDE. 
	For a survey with volume $V_s=10\hhhgpc$ at the median redshift $z=0.5$, a tree-level power spectrum and bispectrum (P+B) analysis out to $k_\text{max}=0.1\hmmpc$ of a nearly 
	unbiased sample of density $\sim 0.01\hhhmpc$ and a biased tracer with $\bar n\sim 10^{-4}\hhhmpc$ and $b_1\sim 3$ can detect these effects with a SNR $\sim$ 
	a few if the DE sound speeds squared are in the range $\hat c_\pm^2 \sim 10^{-2} - 10^{-3}$. 
	Lower values of the sound speeds could be probed with measurements at higher wavenumber, provided that contributions to galaxy clustering are modeled beyond tree-level. 
	Higher values of $\hat c_\pm^2>10^{-2}$ could be accessed with a larger survey volume, with the caveat that the scale-dependent signals decrease with increasing redshift.
	In this regards, weighting the tracers by halo mass could help leverage the scale-dependent bias and improve further the detectability of these effects. While constructing tracer samples that yield ${\rm SNR}>1$ is challenging, it is deemed feasible. 
	For $z\lesssim 0.8$, targeting e.g. luminous red galaxies (LRGs) that preferentially trace rare massive halos of mass $M\gtrsim 10^{13}\,h^{-1}M_\odot$ could help produce
	a biased sample with the above characteristics. 
	By contrast, constructing a nearly unbiased sample with high number density $\bar n\sim 0.01\hhhmpc$ is quite challenging at redshift $z\sim 0.5 - 1$. A deep imaging survey
	could achieve such number densities, but only at the expense of larger (photometric) redshift uncertainties. 
	
	In addition to its influence on the growth of matter perturbations, DDE also acts as a dissipative medium for DM halos if it supports a sound mode.
	Halos moving relative to the DE fluids experience a gravitational drag force analogous to dynamical friction due to the excitation of sound waves, resulting in a transfer of 
	energy from the halo to the DDE component. 
	This introduces both a mass and redshift dependence in the effective growth rate $f\sigma_8$ inferred from (direct or indirect) measurements of halo peculiar velocities.
	This effect is negligible for typical sound speeds, but would become significant if at least one of the DE fluids has an extremely low sound speed 
	($\hat c_\pm^2 \sim 10^{-5}$). 
	In such a limiting case, the peculiar velocities of cluster-size halos would be altered by $\sim 10\%$, with the magnitude of the effect scaling linearly with halo mass. 
	Stacked weak-lensing measurements could provide an avenue to detect the halo-induced DE wakes, although the expected signal is very small and detectable only in Stage-IV
	surveys.
	
	Although a fully self-consistent treatment requires combining these effects with the inclusion of non-linear corrections, our analysis demonstrates that if the DESI results 
	continue to hold, a future full-shape, multi-tracer measurement of galaxy clustering statistics could probe the dynamical nature of dark energy directly at the level of its 
	fluctuations (rather than merely its background evolution) across a broad range of sound speeds.
	
	\section*{Data availability}
	
	The data supporting the findings of this article are publicly available on Github \cite{vandie2026soundmodedata}.
	
	\section*{Acknowledgments}
	
	We thank Gali Eytan and Adi Nusser for discussions, Kwan Chuen Chan and Roman Scoccimarro for comments on an earlier version of this manuscript,
	and the anonymous referee for helpful suggestions. 
	
	\appendix
	
	\section{Covariance matrix for P and B measurements}
	\label{app:covariance}
	
	In the Gaussian approximation to the covariance (obtained upon neglecting off-diagonal elements), 
	the covariance matrix of the power spectrum measurement vector $\vvx_P=(P_{aa},P_{bb},P_{ab})$ evaluated at wavenumber $k$ has the following elements
	\begin{align}
		C_{(aa)(aa)} &= 2\, P_{aa}^2(k)  \\
		C_{(ab)(ab)} &=  P_{ab}^2(k) + P_{aa}(k) \, P_{bb}(k) \nonumber \\
		C_{(aa)(bb)} &= 2\, P_{ab}^2(k) \nonumber \\
		C_{(aa)(ab)} &= 2\, P_{aa}(k)\, P_{ab}(k) \nonumber 
	\end{align}
	with $C_{(bb)(bb)}$ etc. obtained through the permutation $a\leftrightarrow b$. Note that all the power spectra are computed for the same wavenumber
	By contrast, the covariance matrix of the (symmetrized) bispectrum measurement vector $\vvx_B=(B_{aaa},B_{aab},B_{abb},B_{bbb})$ involves power spectra 
	evaluated at 3 different wavenumbers $(k_1,k_2,k_3)$. Namely,
	\begin{align}
		C_{(aaa)(aaa)} &= \Big. 3\, P_{aa}(k_1)\, P_{aa}(k_2)\, P_{aa}(k_3) \\
		C_{(aab)(aab)} &= \frac{1}{3} \Big[P_{aa}(k_1)  P_{aa}(k_2) P_{bb}(k_3) \nonumber \\
		&\qquad + 2 P_{aa}(k_1) P_{ab}(k_2) P_{ab}(k_3)\Big] + \mbox{(2 cyc.)} \nonumber \\
		C_{(aaa)(aab)} &= \Big. P_{aa}(k_1) P_{aa}(k_2) P_{ab}(k_3) + \mbox{(2 cyc.)} \nonumber \\
		C_{(aaa)(abb)} &= \Big. P_{aa}(k_1) P_{ab}(k_2) P_{ab}(k_3) + \mbox{(2 cyc.)} \nonumber \\
		C_{(aaa)(bbb)} &= \Big. 3\ P_{ab}(k_1) P_{ab}(k_2) P_{ab}(k_3) \nonumber \\
		C_{(aab)(abb)} &= \frac{1}{3} \Big[P_{ab}(k_1) P_{ab}(k_2) P_{ab}(k_3) 
		+ P_{aa}(k_1) P_{ab}(k_2) \nonumber \\ &\qquad \times  P_{bb}(k_3) + P_{bb}(k_1) P_{ab}(k_2) P_{aa}(k_3)\Big] \nonumber \\
		&\quad + \mbox{(2 cyc.)} \nonumber
	\end{align}
	while the remaining elements follow from the replacement $a\leftrightarrow b$. 
	The overall factor of 3 in these expressions explains the multiplicative factor of 3 in Eq.~(\ref{eq:FisherB}).
	
	\section{Computation of the friction coefficient}
	\label{app:friction}
	
	\subsection{Finite size effects} 
	\label{app:finite_size}
	
	Since DM halos are extended objects, the point-like approximation for the computation of the friction coefficient may not hold. 
	The correction caused by the finite size of the DM halo enters the Fourier transform of the halo density, 
	\begin{align}
		\rho_h(\omega,\vk) &= \int\!\dd \eta\int\!\dd^3x\,\rho_h(\eta,\vx)\, e^{-i\vk\cdot\vx+i\omega \eta} \\ 
		& = \int\!\dd\eta\, a(\eta)^{-3}M(\eta)\, u(k,\eta)\,e^{-i\vk\cdot\vx_h(\eta)+i\omega \eta} \nonumber \;,
	\end{align}
	through the normalized, spherically symmetric profile $u_h(k,\eta)$ in Fourier space satisfies $u_h(k,\eta)=u_h(-k,\eta)$. 
	For an NFW profile, $u_h(k,\eta)$ is analytic everywhere in the complex plane. 
	Note also that $u_h(k,\eta)$ generally depends on time through the halo concentration $c$, etc. 
	For small $k r_s$ where $r_s$ is the halo virial radius, we have $u_h(k,\eta)\approx 1- \frac{1}{6}\,u_2(\eta)\,(kr_s)^2+ \mathcal{O}((kr_s)^4)$, 
	where the (time-dependent) coefficient $u_2(\eta)=\langle r^2\rangle/r_s^2$ depends on the details of the halo profile. 
	In the subsonic limit $(v_h/\hat c_\pm)\to 0$, only the long-wavelength sound modes with $k r_s\ll 1$ are excited by the motion of the DM halo. 
	As a result, $u_h(k,\eta)\approx 1$ as in the case of a point-like perturber. 
	Therefore, the friction coefficients are insensitive to the finite size effects in the subsonic regime, and thus follow $I_\pm\approx (1/3)(v_h/\hat c_\pm)^3$ 
	so long as $(v_h/\hat c_\pm)\ll 1$.
	
	\begin{figure}
		\centering
		\includegraphics[width=.48\textwidth]{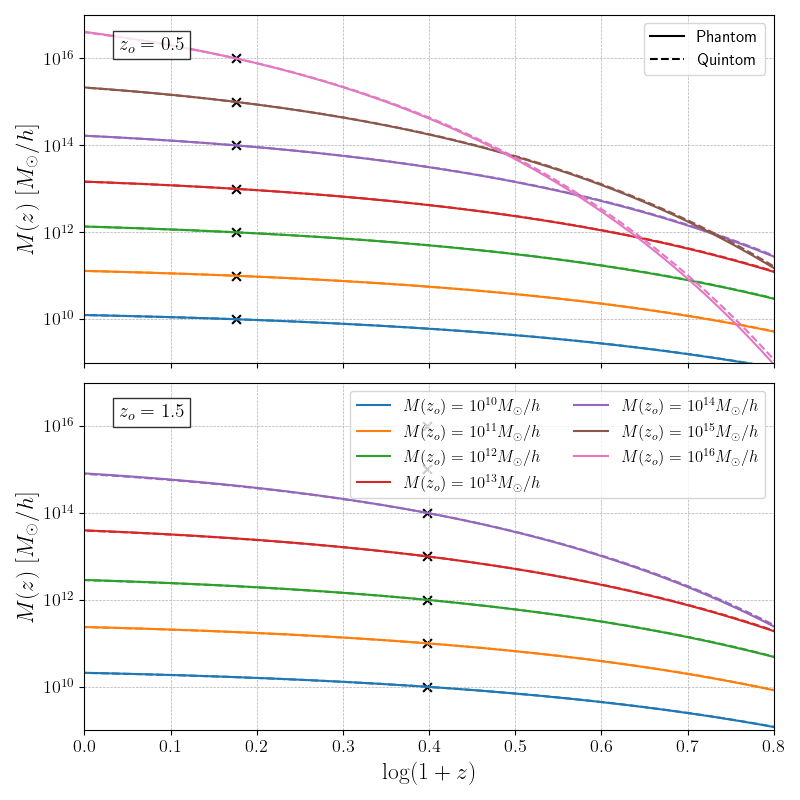}
		\caption{Evolution of DM halo masses, according to the model described in the main text. 
			Halo masses are normalized at redshift $z_o=0.5$ (top panel) and $1.5$ (bottom panel).
			$M(z_o) = 10^{15} \hmsun$ and $M(z_o) = 10^{16} \hmsun$ are not shown in the bottom panel because the model does not work accurately at high masses, 
			which is the reason why we limit ourselves to $M(z_o) = 10^{14}\hmsun$ for $z_o\geq 1$.}
		\label{fig:hmassevol}
	\end{figure}
	
	\subsection{Mass accretion history} 
	\label{sec:mass_accretion}
	
	We include the average mass accretion history of DM halos following the approach of \cite{Correa:2014xma} based on the extended Press-Schechter (EPS) formalism 
	(see also \cite{Reed:2003sq,Genel2010,Bosch2014}). 
	Concretely, we assume a redshift-dependent halo mass 
	\begin{equation}
		M(z) = M_0 (1 + z)^{\alpha} e^{\beta z} 
	\end{equation}
	where $M_0$ is the present-day halo mass. 
	The mass accretion over cosmic time is governed by $\alpha$ and $\beta$, which determine the low-$z$ (late-time) and  high-$z$ (early-time) growth of the halo mass, respectively. 
	They are given by
	\begin{equation}
		\alpha = a_c \, f_c(M_0)\;, \quad \beta = -f_c(M_0) \;,
	\end{equation}
	where the parameter
	\begin{equation}
		a_c = \delta_c \sqrt{\frac{2}{\pi}} \dv{D}{z} \Big|_{z=0} + 1\;,
	\end{equation}
	depends on the critical overdensity $\delta_c\approx 1.686$ for (spherical) collapse in the EPS formalism \cite{gunn/gott:1972,EPS}. 
	The sensitivity of the growth rate to the background evolution results in somewhat different values of $a_c$ -- namely, $a_c\approx 0.12$, 0.29, and 0.78 for the Quintom, 
	$\Lambda$CDM and Phantom models -- and consequently to modest differences in their mass accretion histories.
	Note that, while we should have used the growth rate on sub-Jeans scales, we have ignored this complication because the scale-dependence in the growth rate $D(z,k)$ of matter 
	fluctuations is very mild for the models considered here (see Section~\S\ref{sec:scalekbias}).
	
	Furthermore, the dependence on the statistical properties of the matter density fluctuations enters the (positive definite) quantity
	\begin{equation}
		f_c(M_0) = \left[S\!\left( \frac{M_0}{q} \right) - S(M_0) \right]^{-1/2}\;,
	\end{equation}
	through the variance $S(M) = \sigma^2(M)$ of the density field smoothed on the mass scale $M$ (with a top-hat filter). The parameter 
	\begin{equation}
		q = 4.137\tilde{z_f}^{-0.9476}
	\end{equation}
	is related to the average formation redshift 
	\begin{equation}
		\tilde{z_f} = -0.0064(\log_{10} M_0)^2 + 0.0237\log_{10}M_0 + 1.8837
	\end{equation}
	of the DM halo~\footnote{In contrast to \cite{Correa:2014xma}, we use the relation $q = (4.137/\tilde{z_f})^{0.9476}$, which turns out to fit their Fig.~2 better. 
		We attribute this difference to a typo. Moreover, the numerical coefficients in the $q$ - $z$ relation were originally computed for a WMAP5 cosmology, 
		but the authors indicate that it also works for a Planck cosmology. We therefore assume the relation holds in DDE cosmologies as well.}.
	
	Finally, $\beta$ is negative, which reflects the hierarchical growth of halo mass in matter domination. 
	Likewise, the low-$z$ power-law scaling $(1 + z)^{\alpha}$ is caused by the suppression of the growth rate of matter density perturbations as the expansion accelerates. 
	The dependence of $\alpha$ on the parameter $a_c$ links this effect to $\dot D$, whereas $f_c(M_0)$ introduces a dependence on halo mass through the variance of the matter density field.
	
	As seen in Fig.\ref{fig:hmassevol}, the model loses accuracy for high halo masses, as the corresponding mass growth curves intersect at sufficiently high redshift. 
	However, since massive halos are rare and the friction induced by DDE becomes relevant at redshift $z\sim 2$, this remains a relatively minor issue. 
	
	\bibliography{references}
	
\end{document}